\UseRawInputEncoding

\documentclass[twocolumn]{aastex631}
\usepackage{multirow}
\usepackage{graphicx}   
\usepackage{amsmath}    
\usepackage{amssymb}    
\usepackage{threeparttable}
\usepackage{array}
\usepackage[figuresright]{rotating}
\usepackage{ulem}
\usepackage{hyperref}
\usepackage{cleveref}
\newcommand{\revise}[1]{{#1}}
\newcommand{\newrevise}[1]{{#1}}

\received{February 23, 2022}
\revised{July 30, 2022}
\accepted{September 14, 2022}

\submitjournal{ApJ}

\shorttitle{Optical Variability of ELGs}
\shortauthors{Lin, Zheng, et al.}
\graphicspath{{./}{figures/}}

\begin{document}

\title{On the Origin of the Strong Optical Variability of Emission-line Galaxies}

\author{Ruqiu Lin}
\affiliation{Key Laboratory for Research in Galaxies and Cosmology, Shanghai Astronomical Observatory, Chinese Academy of Sciences, 80 Nandan Road, Shanghai 200030, People's Republic of China}
\affiliation{School of Astronomy and Space Sciences, University of Chinese Academy of Sciences, No. 19A Yuquan Road, Beijing 100049, People's Republic of China}

\author{Zhen-Ya Zheng*}
\affiliation{Key Laboratory for Research in Galaxies and Cosmology, Shanghai Astronomical Observatory, Chinese Academy of Sciences, 80 Nandan Road, Shanghai 200030, People's Republic of China}
\correspondingauthor{Zhen-Ya Zheng}
\email{zhengzy@shao.ac.cn}

\author{Weida Hu}
\affiliation{CAS Key Laboratory for Research in Galaxies and Cosmology, Department of Astronomy, University of Science and Technology of China,
Hefei, Anhui 230026, People's Republic of China}

\author{Chunyan Jiang}
\affiliation{Key Laboratory for Research in Galaxies and Cosmology, Shanghai Astronomical Observatory, Chinese Academy of Sciences, 80 Nandan Road, Shanghai 200030, People's Republic of China}

\author{Xiang Pan}
\affiliation{Key Laboratory for Polar Science, MNR, Polar Research Institute of China, 451 Jinqiao Road, Shanghai, 200136, People's Republic of China}

\author{Chenwei Yang}
\affiliation{Key Laboratory for Polar Science, MNR, Polar Research Institute of China, 451 Jinqiao Road, Shanghai, 200136, People's Republic of China}

\author{Fang-ting Yuan}
\affiliation{Key Laboratory for Research in Galaxies and Cosmology, Shanghai Astronomical Observatory, Chinese Academy of Sciences, 80 Nandan Road, Shanghai 200030, People's Republic of China}

\author{Rahna P.T.}
\affiliation{Key Laboratory for Research in Galaxies and Cosmology, Shanghai Astronomical Observatory, Chinese Academy of Sciences, 80 Nandan Road, Shanghai 200030, People's Republic of China}

\author{Jian-Guo Wang}
\affiliation{Yunnan Observatories (YNAO), Chinese Academy of Sciences, Kunming 650216, Peopleʼs Republic of China}
\affiliation{Key Laboratory for the Structure and Evolution of Celestial Objects,CAS, Kunming, 650216, Peopleʼs Republic of China}

\author{Yibo Wang}
\affiliation{CAS Key Laboratory for Research in Galaxies and Cosmology, Department of Astronomy, University of Science and Technology of China,
Hefei, Anhui 230026, People's Republic of China}

\author{Ning Jiang}
\affiliation{CAS Key Laboratory for Research in Galaxies and Cosmology, Department of Astronomy, University of Science and Technology of China,
Hefei, Anhui 230026, People's Republic of China}

\author{Shuairu Zhu}
\affiliation{Key Laboratory for Research in Galaxies and Cosmology, Shanghai Astronomical Observatory, Chinese Academy of Sciences, 80 Nandan Road, Shanghai 200030, People's Republic of China}
\affiliation{School of Astronomy and Space Sciences, University of Chinese Academy of Sciences, No. 19A Yuquan Road, Beijing 100049, People's Republic of China}








\begin{abstract}
Emission-line galaxies (ELGs) are crucial in understanding the formation and evolution of galaxies, while little is known about their variability.
Here we report the study on the optical variability of a sample of ELGs selected in the COSMOS field, which has narrow-band observations in two epochs separated by $\gtrsim$ 12 years.
This sample was observed with Suprime-Cam (SC) and Hyper Suprime-Cam (HSC) on the $Subaru$ telescope in NB816 and $i'/i$ bands, respectively. After carefully removing the wing effect of a narrow-band filter, we check the optical variability in a sample of 181 spectroscopically confirmed ELGs. We find that 0 (0/68) H$\alpha$ emitters, 11.9\% (5/42)  [O\,{\sc iii}] emitters, and \revise{0 (0/71)} [O\,{\sc ii}] emitters show significant variability ($|\Delta m_\textrm{NB}| \geq 3\,\sigma_{\Delta m_\textrm{NB,AGN}} = \revise{0.20}\, \textrm{mag}$) in the two-epoch narrow-band observations. We investigate the presence of active galactic nucleus (AGN) in this variable ELG (var-ELG) sample with three methods, including X-ray luminosity, mid-infrared activity, and radio-excess. We find zero bright AGN in this var-ELG sample, but cannot rule out the contribution from faint AGN. We find that SNe could also dominate the variability of the var-ELG sample. The merger morphology shown in the HST/F814W images of all the var-ELG sample is in agreement with the enhancement of star formation, i.e., the SNe activity. 



\end{abstract}



\section{Introduction} 
\label{sec:intro}

Galaxies with strong emission-lines in their spectra are known as emission-line galaxies (ELGs), which are typically less massive, less dusty, and more efficient in star-forming than normal galaxies~\citep[e.g.,][]{griffiths_emission_2021}.
Studying ELGs improves our understanding of the formation and evolution of distant galaxies, galaxy clusters, and cosmology~\citep[][]{benitez2014}. 
ELGs have been largely surveyed by various methods, e.g., spectroscopic surveys~\citep[e.g., SDSS,][]{York2000}, grism surveys~\citep[e.g.][]{momcheva2016}, and narrow-band imaging surveys ~\citep[NB surveys, e.g.][and references therein]{Geach2008,Sobral2015}.
Among those methods, the narrow-band imaging is the most efficient survey method for ELGs, as it covers the corresponding emission-line (e.g., H$\alpha$,  [O\,{\sc iii}], [O\,{\sc ii}] or Ly$\alpha$) of ELGs in a narrow redshift range.


In galaxies, the radiation of emission-lines is mainly contributed by star-formation, activities of central black holes, and shock excitation~\citep[see][]{kewley2019}.
Nebular emission-lines, particularly H$\alpha$, trace star-formation rates (SFRs) in ELGs~\citep{Kennicutt1998}, because the H$\alpha$ emission is closely related to the photoionization driven by UV radiation from young and blue stars. \revise{Beside}, other emission lines such as [O\,{\sc iii}] and [O\,{\sc ii}] can be SFRs indicators as well~\citep[see][]{Madau2014}.
The radiative shock driven by supernova would produce emission lines 
 in the cooling of shock-heated gas~\citep[e.g.,][]{cox1972, sutherland2017}.
Meanwhile, Active galactic nuclei (AGN) often show strong emission-lines in their spectra, which are marked as the evidence of the central super massive black holes (\citealt[also see]{krolik1978,ferland1979,kallman1982}~\citealt{kewley2019}). 

In the past two decades, a large number of narrow-band surveys 
search and study ELGs at various redshifts from $z\sim0$ to 8~\citep[][]{Ouchi2020}, which makes it possible to construct the star-forming history of the universe~\citep[e.g., see][]{Madau2014}.
For example, in the well-known COSMOS field~\citep{Scoville2007}, ELGs from NB surveys provide clues to star formation in the redshift range from $z$ $\sim$ 0.05 to $z$ $\sim$ 1.47~\citep{Geach2008, Hayashi2018}.  
However, there is little work focusing on the optical variability of ELGs.

The optical variability of ELGs can be explained by three origins: (1) AGN activities. Variability is a basic phenomenon of AGN, which could be explained by various accretion disk and wind models~\citep[e.g.,][]{Ulrich1997, Choi2014, Heinis2016, Caplar2017, Cai2018, De-Cicco2019}. In addition, changing-look AGN could produce broad emission-lines and high ionization lines in about several years ~\citep{LaMassa2015}. (2) supernovae (SNe) explosions. SNe explosions would lighten the host galaxy in a time scale of months. Furthermore,  the regions heated by SNe shocks significantly influence the radiation of emission-lines, which evolve during the interaction between SNe and no-shock regions surrounded~\citep[e.g., see][]{tartaglia2020}. (3) stellar tidal disruption events (TDEs). TDEs in a gas-rich environment could produce strong H$\alpha$ and [O\,{\sc iii}] emissions with no significant change in the continuum~\citep{Yang2013}. 
So the question is, which process dominates the strong optical variability of ELGs?
The goal of this paper is to reveal the origin of the optical variability of ELGs. We use the narrow-band and broad-band imaging data taken in the COSMOS field in two epochs separated by $\gtrsim$12 years to test the optical variability of ELGs. 
We then analyze the presence of AGN in the spectroscopically confirmed ELGs which show variability in the narrow-band data. 

This paper is organized as follows. In \S~\ref{sec:2} we introduce the sample selection of ELGs and the narrow-band variability test method.
The results of the variability test, the morphology, and the AGN fraction of our sample of strong variable ELGs are presented in \S~\ref{sec:results}. In \S~\ref{sec:discussion} we discuss the AGN fraction \revise{in the whole ELG sample and possible origin of strong variability of our variable ELG sample.} The result is concluded in \S~\ref{sec:conclusion}.
Throughout this paper, we assume the cosmological parameters of ${\rm H0 = 70\, km\,s^{-1}\,Mpc^{-1},\, \Omega_m = 0.3,\, and\ \Omega_\Lambda = 0.7}$.

\section{The Sample of ELGs and The Variability Test Method}
\label{sec:2}



\begin{figure}
    \centering
    \includegraphics[width=0.49\textwidth]{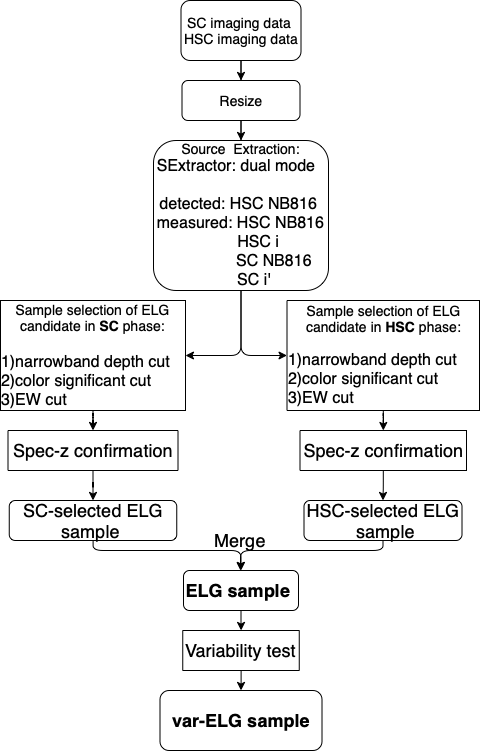}
    \caption{Processing flow on the sample selection of ELGs and variable ELGs.}
    \label{fig:process}
\end{figure}


\subsection{Data Description}
\label{sec:data_description}

The ELG sample in this work are selected with the narrow-band NB816 and broad-band imaging data in the COSMOS field taken with Suprime-Cam (SC)~\citep[]{Miyazaki2002} and Hyper Suprime-Cam (HSC)~\citep[]{miyazaki2018} on the $Subaru$ 8.2 meters telescope. These NB816 images, observed with SC in 2004-2005 and HSC in 2016 and 2019, allow us to analyze the two-epoch narrow-band variability of ELGs.  The summary of the two-epoch observations is listed in Table~\ref{tab:filter}. 
The processing flow on the sample selection of ELGs and the variability test is shown in Figure~\ref{fig:process}.

\begin{table*}
    \centering
    \caption{Summary of the two-epoch broad-band and narrow-band observations with SC and HSC on the {\it Subaru} telescope}
    \label{tab:filter}
    \setlength{\tabcolsep}{0.70cm}{
    \begin{tabular}{lccccc}\hline \hline
    Instrument  & \multicolumn{2}{c}{SC}  &  & \multicolumn{2}{c}{HSC} \\  
    \cline{2-3}  \cline{5-6} 
    Observing Date$^a$ & 2004 &   2004-2005 
    & & 2016-2020 & 2016, 2019
    \\  
    \cline{2-3} \cline{5-6} 
    Filter                & $i'$     & NB816   & & $i$      & NB816 \\
    $\lambda_c$ (\AA)     & 7641     & 8151    & & 7711     & 8177  \\
    $\Delta \lambda$ (\AA)& 1497     & 117     & & 1574     & 113   \\
    Depth (mag)           & 26.2     & 25.7    & & 26.9     & 26.0  \\
    Saturation (mag)      & 20.0     & 16.9    & & 18.6     & 16.8  \\ 
    Seeing(arcsec)$^b$    & 0.7-1.2  & 0.7-1.2 & & 0.6-0.7  & 0.6-0.8 \\
    \hline 
    \multicolumn{6}{l}{$^a$ See Table~\ref{tab:observing_date} in details.} \\
    \multicolumn{6}{l}{$^b$ Seeing values are FWHMs within 1$\sigma$ errors of unsaturated point sources from the stacked images.} \\
    \multicolumn{6}{l}{References of the observations: \citet[]{Taniguchi2007}, ~\citet[]{Capak2007}, ~\citet[]{Aihara2021}, ~\citet{Hayashi2018}.} \\
    \end{tabular}}
\end{table*}

In 2004-2005 Subaru took an imaging survey with SC in the COSMOS field with panchromatic bands~\citep[]{Taniguchi2007,Capak2007} including NB816 and $i'$. These images are available from IRSA\footnote{https://irsa.ipac.caltech.edu/} data \revise{server}. SC worked in 2014-2017 with a shared-risk mode with HSC, and was fully decommissioned in 2017. 

Since 2014, Subaru had initiated the Hyper Suprime-Cam Subaru Strategic Survey Program~\citep[HSC-SSP,][]{Aihara2018}, which is a multi-band imaging survey with wide, deep, and/or ultra-deep exposures in many famous fields with its new imager HSC. The COSMOS field is covered by HSC-SSP deep and ultra-deep surveys. The HSC $i$-band image used here, archived from HSC-SSP public data release 3~\citep[PDR3,][]{Aihara2021}, is a combination of all observations taken from February 2016 to January 2020.

\subsection{Source Extraction and Photometry}
\label{sec:source_extraction}
We preform source extraction and photometry using SE{\sc xtractor}~\citep{Bertin1996} in dual-mode, which detect sources from the narrow-band NB816 image and measure colors from the broad-band ($i$/$i'$) image. Since the HSC-NB816 band is  deeper and with better image quality than the SC-NB816 band (see Table~\ref{tab:filter}), we choose HSC-NB816 as the detection image.  We also test the results with HSC-{\it i} as the detection image, which has little affection on the final result. 
Since the automatic aperture magnitude MAG\_AUTO is greatly affected by source blending and image seeing, 
we choose the fixed aperture magnitude rather than the automatic aperture magnitude in the following calculation. 
To check the seeing (PSF) effect, fixed aperture magnitudes are derived in both 3\arcsec\,and 6\arcsec\,diameters. The narrow-band variability tests with the two apertures have similar results, while a small fraction of ELGs show large uncertainties in their broad-band.   
This is because we choose the narrow-band image as the detection image, so a larger aperture would cause more contamination of nearby objects in the deeper broad-band images. Therefore, we choose the aperture magnitude in a 3\arcsec\,aperture in the following analysis.

\revise{Since {\sc SExtractor} would underestimate the photometric errors \citep[see][]{Laigle2016,bielby2012}, we correct this effect by comparing the average 1-$\sigma$ error from faint (non-saturated) objects detected by {\sc SExtractor} and the standard deviation of fluxes obtained from empty background apertures. We randomly place $\sim$40,000 apertures with same 3\arcsec\ diameters in the blank sky region to obtain the standard deviation of background fluctuations. Then the photometric errors from {\sc SExtractor} are corrected by  multiplying the ratio between the standard deviation of background fluctuations and the average 1-$\sigma$ error from  {\sc SExtractor}. The correction factors are $f_{SC-i'}$=1.9, $f_{SC-NB}$=1.7, $f_{HSC-i}$=3.2, and $f_{SC-NB}$=2.4 for SC-{\it i'}, SC-NB816, HSC-{\it i}, and HSC-NB816, respectively. After this correction, the differences of  photometric errors between our catalog and the COSMOS2015~\citep[]{Laigle2016} catalog for SC-selected ELGs in SC-NB and SC-{\it i'} bands are below 0.04 and 0.08 mag, respectively.}



There are \revise{1,187,790} sources detected in HSC-NB816 band in the COSMOS field.
After excluding saturated sources and bad areas marked in COSMOS2015~\citep[]{Laigle2016}, \revise{653,524} sources are left. We then collect their aperture magnitudes in HSC-$i$, SC-$i'$, and SC-NB816 bands, for the following ELG selection and variability test. The photometric zero-point (ZP) is derived by a comparison of point-source photometry (3\arcsec\,diameters) between our catalog and the COSMOS2015 catalog in the corresponding bands.

\subsection{ELG Candidate Selection and Redshift Validation}
\label{sec:sample} 

\begin{figure*}
    \includegraphics[width=1.08\columnwidth]{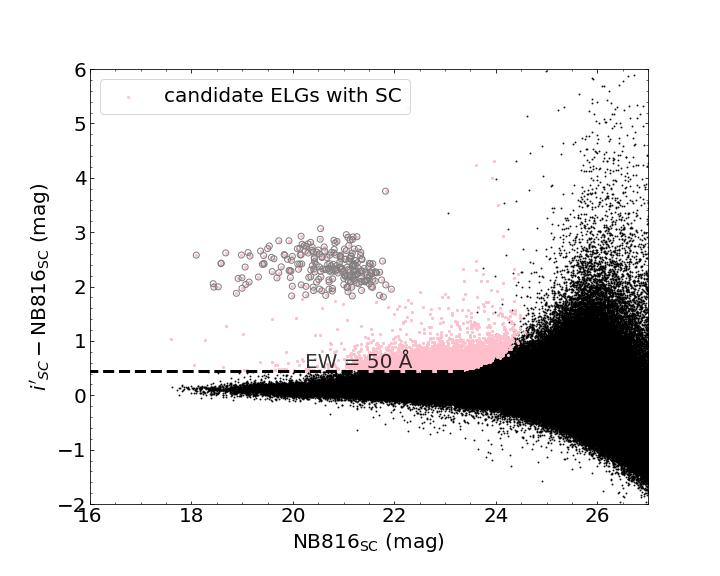}
    \includegraphics[width=1.08\columnwidth]{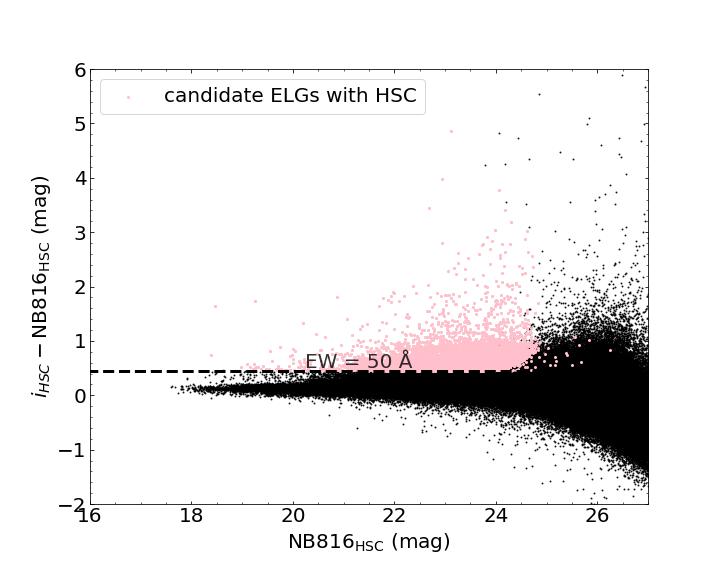}
    \caption{Color-magnitude diagram for selecting candidate ELGs based on NB816 and {\it i'/i} magnitude with SC ({\it left panel}) and HSC ({\it right panel}), respectively. The black dots represent all non-saturated sources. Pink dots represent ELG candidates selected. The black dashed lines mark the magnitude excess for an observed line equivalent width (EW) cut of 28 \AA. \revise{Further more, the gray circles marked in the left panel are the saturated sources in the SC-{\it i'} band image because of the much lower saturation magnitude in the SC-{\it i'} band ({\it i'}$_\textrm{SC, satur} = 20.0$) than the HSC-{\it i} band ({\it i}$_\textrm{HSC, satur} = 18.6$).
    }}
    \label{fig:ELG_candidate_selection}
\end{figure*}

We use the narrow-band imaging technique to effectively select ELG candidates in each epoch. The selection criteria of the ELG candidates include a cut on the narrow-band magnitude, a significant excess of the broad-band over narrow-band color, and a cut on the equivalent-width of emission-lines.

Firstly, sources are required with a $\geq$ 5 $\sigma$ detection in the narrow-band NB816. 
We then apply a color significance cut, $\Sigma\geq$  3, to exclude cases of false candidate ELGs caused by photometric scatters. The color significance $\Sigma$ is defined as:
\begin{equation}
    \centering
    \label{equ:Sigma}
    \Sigma = \frac{BB-NB}{\sqrt{\sigma_{BB}^2 + \sigma_{NB}^2}},
\end{equation} 
where NB and $\sigma_{NB}$ (BB and $\sigma_{BB}$) are the 3\arcsec\,diameter aperture magnitudes and their errors in NB816 ($i'/i$ band), respectively.




An observed equivalent width (EW$_{obs}$) cut of \revise{28} \AA\ is applied to ensure that photons from galaxies in the narrow-band filter are dominated by their emission-line instead of their continuum. The emission-line flux $F_{line}$, continuum flux density $f_c$, and the EW$_{obs}$ of the emission-line are defined as:
\begin{eqnarray}
    \label{equ:flux_line}
    F_{line} & = & \Delta\lambda_{NB}\frac{f_{NB}-f_{BB}}{1-(\Delta\lambda_{NB}/\Delta\lambda_{BB})}, \\
    f_c & = & \frac{f_{BB}-f_{NB}(\Delta\lambda_{NB}/\Delta\lambda_{BB})}{1-(\Delta\lambda_{NB}/\Delta\lambda_{BB})}, \\
    EW_{obs} & = & \frac{F_{line}}{f_c} = \Delta\lambda_{NB}\frac{f_{NB}-f_{BB}}{f_{BB}-f_{NB}(\Delta\lambda_{NB}/\Delta\lambda_{BB})},
\end{eqnarray}
where $f_{NB}$ and $f_{BB}$ are the flux density in the narrow-band and in the broad-band, respectively. We adopt the narrow-band bandwidths ($\Delta \lambda_{NB}$) of 113 \AA, and the broad-band width of ($\Delta \lambda_{BB}$) of 1574 \AA. 
The corresponding criterion for EW$_{\textrm{obs}} >$ \revise{28} \AA\,is (BB - NB) $>$ 0.35 ($f_{NB}/f_{BB} > 1.38$) with assumptions of a flat continuum spectrum and a same central wavelength for NB and BB filters. However, because the central wavelength of the NB filter is red-ward from the center of the BB filter, non-flat continuum falling in BB filter will have a color offset of (BB - NB). This color offset zeropoint is calibrated as $\sim$ 0.1 mag with the matched point sources which show continuum dominated in the BB and NB filters. 
Considering the color offset of 0.1 mag, we apply (BB - NB) $\geq$ 0.45 mag in the sample selection (Figure~\ref{fig:ELG_candidate_selection}). 
With the above three criteria, we select two ELG samples with Subaru-SC and Subaru-HSC, which contain \revise{4,975} and \revise{6,347} ELG candidates, respectively. 

\begin{figure*}
    {\centering
    \includegraphics[width=1.0\columnwidth]{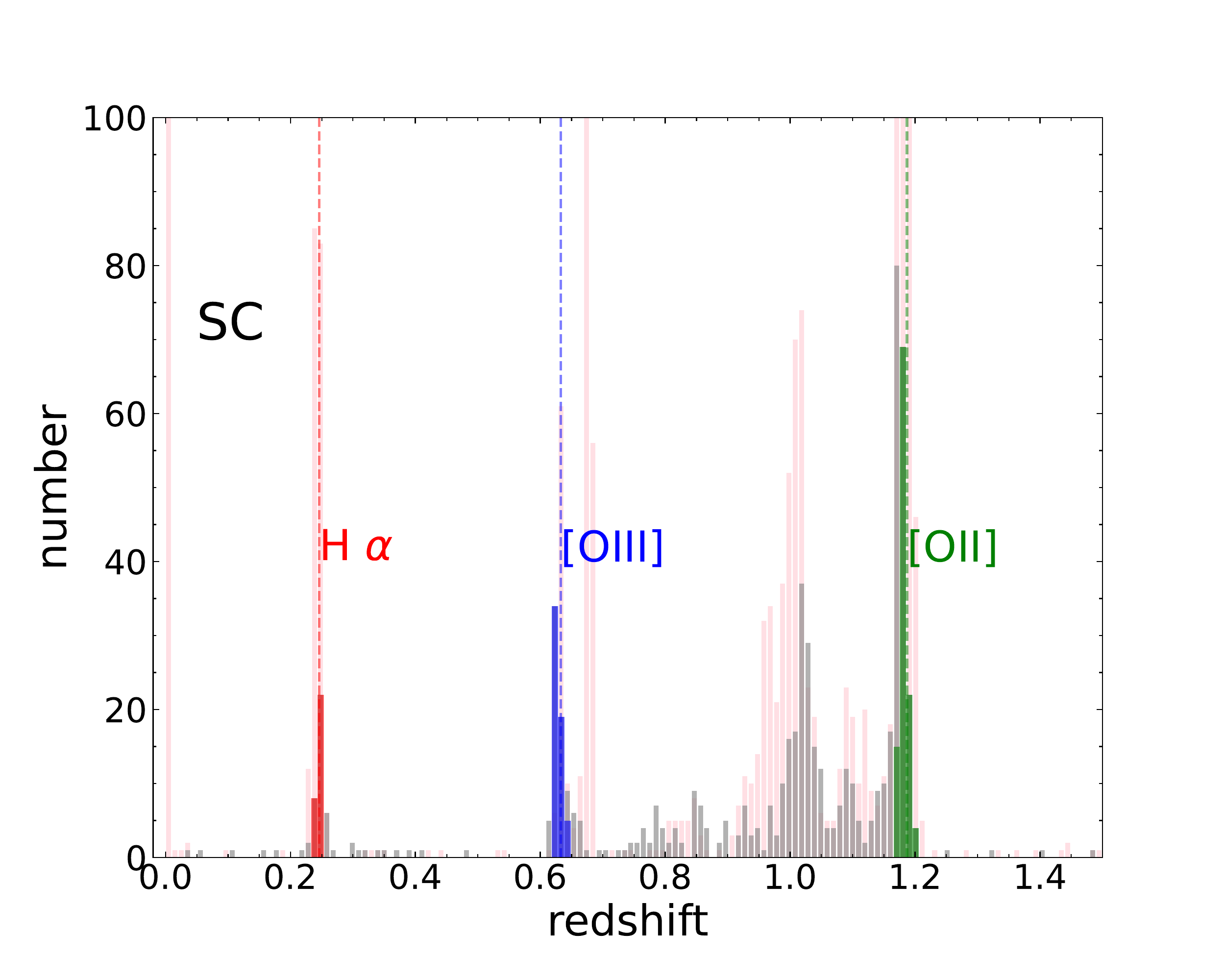}
    \includegraphics[width=1.0\columnwidth]{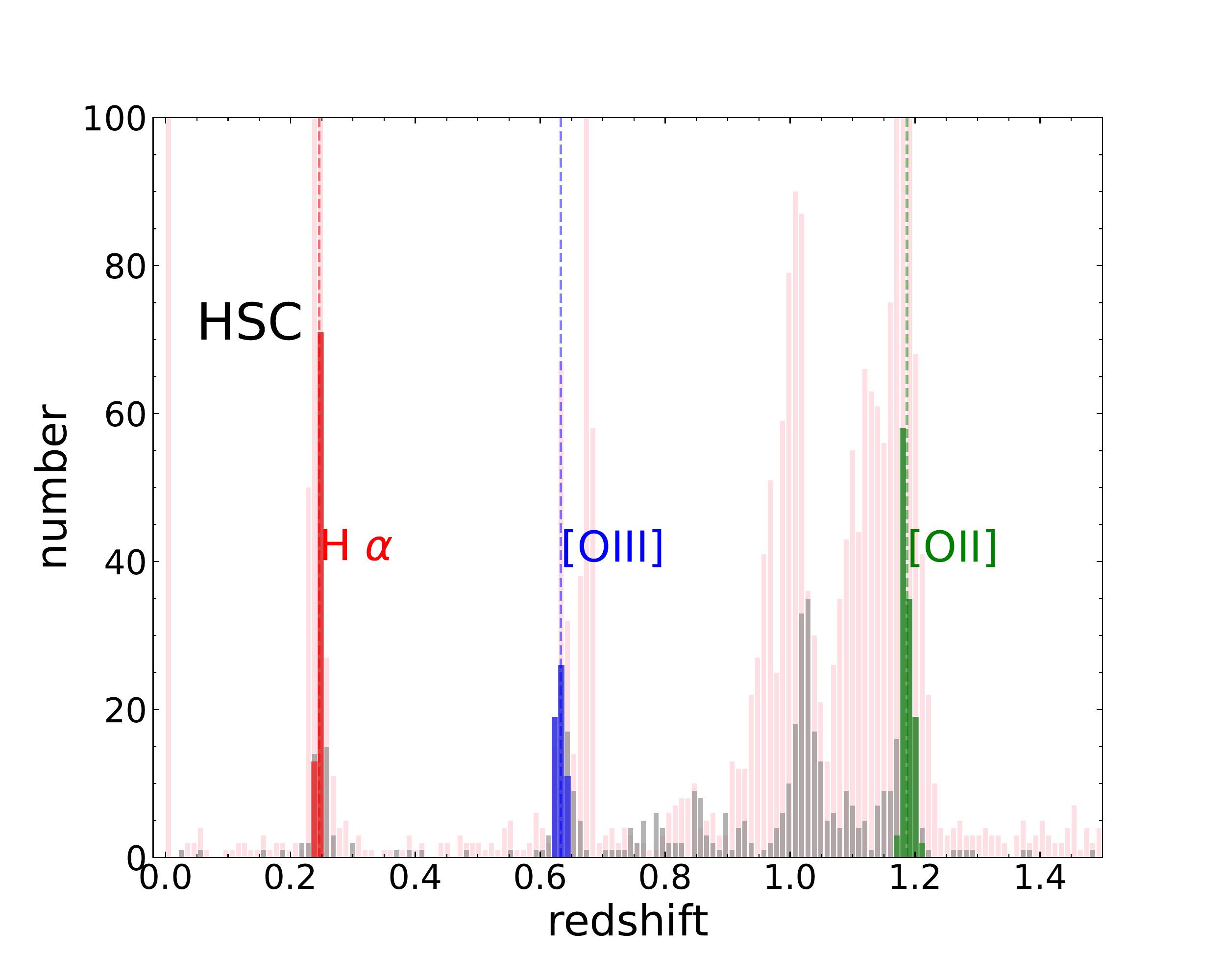}
    \caption{
     Redshift distributions of ELG candidates selected with SC ({\it left panel}) and HSC ({\it right panel}). The distributions of photometric \newrevise{redshifts (COSMOS2015, \citealt{Laigle2016})} and spectroscopic redshifts \newrevise{(see \S~\ref{sec:sample}, the 4th paragraph for details)} of ELG candidates are marked in pink and grey histograms, respectively. The spectroscopically confirmed H$\alpha$, [O\,{\sc iii}] and [O\,{\sc ii}] emitters are marked in red, blue and green colors, respectively.}
    \label{fig:z-distirbution}}
\end{figure*}

\begin{table*}
    \centering
	\caption{Statistics of the ELG sample}
	\label{tab:ELGs}
	\setlength{\tabcolsep}{0.03cm}{
    \begin{tabular}{ccccccccc} \hline\hline
    \multirow{3}{*}{Line} &
    \multirow{3}{*}{Spec-{\it z} range} &
    \multirow{3}{*}{\#(ELGs)} &
        \multirow{3}{*}{\#(var-ELGs)} &
    \multicolumn{5}{c}{\#(AGN) in ELGs} \\ 
     \cline{5-9} &  &  & & \begin{tabular}[c]{@{}c@{}}X-ray\\ 
    selected\end{tabular} & \begin{tabular}[c]{@{}c@{}}MIR\\ 
    selected\end{tabular} & \begin{tabular}[c]{@{}c@{}}Radio\\
    selected\end{tabular} & \begin{tabular}[c]{@{}c@{}}\revise{BPT}\\
    \revise{selected}\end{tabular} &  \begin{tabular}[c]{@{}c@{}}\#(AGN)\\ 
    \end{tabular} 
    \\ 
    \hline
    H$\alpha$           & 0.238 - 0.250     &  68 & 0  &  1 $(0^a)$ & 0   &0   & \revise{2/42}$^e$ & \revise{3}     \\
    & & & & [41.5]$^b$ & [41.1,41.3,42.0,42.0]$^c$ & [34.0,35.1]$^d$ &  & \\
    {[}O{\sc iii}{]}    & 0.623 - 0.639     &  \revise{42} & \revise{5} &  0         & 0   &0   & \revise{--} & 0     \\
    & & & & [42.2]$^b$ & [42.1,42.3,43.0,42.9]$^c$ & [34.9,36.0]$^d$ &  & \\
    {[}O{\sc ii}{]}     & 1.181 - 1.202     &  \revise{71} & \revise{0} &  \revise{5 $(5^a)$} & 11  &\revise{2}   & \revise{--} & \revise{14}    \\
    & & & & [42.9]$^b$ & [42.8,42.9,43.7,43.6]$^c$ & [35.6,36.7]$^d$ &  & \\
    \hline
    \end{tabular}
    }
    \begin{tablenotes}
    \footnotesize
    \item $^a$ Selection criterion of X-ray bright AGN: $\rm L_X \geqq 10^{43}\ erg\cdot s^{-1}$. 
    \item $^b$ \revise{The 5-$\sigma$ luminosity (log$L$) detection limit in the X-ray 0.5-10 keV band. }
    \item $^c$ \revise{The 5-$\sigma$ luminosity (log$\nu L_\nu$) detection limits in the 3.6, 4.5, 5.8, and 8.0 $\mu m$ bands. }
    \item $^d$ \revise{The 5-$\sigma$ luminosity (log$\nu L_\nu$) detection limits in the 3 and 1.4 GHz bands.}
    \item $^e$ \revise{Only 42 out of 68 H$\alpha$ emitters were covered by the $z$COSMOS spectroscopic survey, and the BPT diagnostic was applied to this sub-sample.}
    \end{tablenotes}
\end{table*}

The enormous spectroscopic resources in the COSMOS field are used to validate the redshifts of ELG candidates. Spectroscopic redshifts are available in HSC-SSP public data release 3 (HSC-SSP PDR3), which includes several surveys such as $z$COSMOS DR3~\citep{Lilly2009}, UDSz~\citep{Bradshaw2013, McLure2013}, 3D-HST~\citep{Skelton2014, momcheva2016}, FMOS-COSMOS~\citep{Silverman2015}, VVDS~\citep{Le-Fevre2013}, VIPERS PDR1~\citep{Garilli2014}, SDSS DR16~\citep{Ahumada2020}, SDSS QSO DR14~\citep{Paris2018},  GAMA DR2~\citep{Liske2015}, WiggleZ DR1~\citep{Drinkwater2010}, DEEP2 DR4 ~\citep{Davis2003,Newman2013}, DEEP3~\citep{Cooper2011,Cooper2012}, PRIMUS DR1~\citep{Coil2011,Cool2013},2dFGRS~\citep{Colless2003}, 6dFGRS~\citep{Jones2004, Jones2009}, C3R2 DR2~\citep{Masters2017, Masters2019}, DEIMOS 10k sample~\citep{Hasinger2018}, LEGA-C DR2~\citep{Straatman2018}, and VANDELS DR1~\citep{Pentericci2018}. 
We then cross-match the ELG candidate catalogs with these spectroscopic redshift catalogs with a match radius of 0\farcs5. The matched samples include \revise{647} and \revise{648} sources from SC- and HSC-selected ELG candidates, respectively. \newrevise{Only 144 candidates are in both SC- and HSC-selected candidate ELG catalogs.}

We then use spectroscopic redshifts to select true H$\alpha$, [O\,{\sc iii}] and [O\,{\sc ii}] emitters from ELG candidates (see Figure~\ref{fig:z-distirbution}). 
The expected redshifts of H$\alpha$, [O\,{\sc iii}] and [O\,{\sc ii}] emitters are approximately 0.24, 0.63 and 1.19 for the NB816 selected ELGs, respectively. Figure~\ref{fig:z-distirbution} presents the redshift distribution of ELG candidates selected with NB816 and i bands. 
By checking the profiles of the two NB816 filters, we constrain the observed wavelength of ELGs' emission-lines $\lambda_{obs}$ within 8127-8205 \AA\ to rule out the wing effect of the narrow-band filter (see \S~\ref{sec:Variability} in detail). Therefore, the corresponding redshift ranges are 0.238$<z<$0.250, 0.623$<z<$0.639, and 1.181$<z<$1.202 for H$\alpha$, [O\,{\sc iii}] and [O\,{\sc ii}] emitters, respectively (also listed in Table~\ref{tab:ELGs}). \newrevise{Only $\sim$ 16\% (27\%) of 647 (648) SC-selected (HSC-selected) candidate ELGs are confirmed as true sample of H$\alpha$, [O\,{\sc iii}] and [O\,{\sc ii}] ELGs by their spectroscopic redshifts. Most of the remaining candidates are actually Balmer-break galaxies or 4000A-break galaxies located in the redshift range of 0.7--1.1 (see Figure~\ref{fig:z-distirbution}, the grey histograms for the distributions of spectroscopic redshifts). We should address that with extra criteria such as photo-z range and/or broadband color-color selection, the contamination fraction in the ELG candidates would be significantly decreased  \citep[e.g.,][]{Khostovan2020}. In this work, since we need the accurate redshift information, we choose to match the spectroscopic catalogs directly and do not need to worry about the contamination.}

The spectroscopically confirmed ELGs selected with SC and HSC are then merged together with TOPCAT~\citep{Taylor2005} as the final ELG sample. In a total number of \revise{181} ELGs, \revise{93} are selected in both SC and HSC, \revise{8} are selected in SC only, and \revise{80} are selected in HSC only. Sorted by their redshifts, we obtain 26 H$\alpha$, \revise{33} [O\,{\sc iii}] and \revise{42} [O\,{\sc ii}] emitters selected in SC, and 67 H$\alpha$, \revise{36} [O\,{\sc iii}] and 70 [O\,{\sc ii}] emitters selected in HSC. 
 Finally, the combined ELG sample includes 68 H$\alpha$, \revise{42} [O\,{\sc iii}], and \revise{71} [O\,{\sc ii}] emitters (see also Table~\ref{tab:ELGs}).


\subsection{Optical Variability Test} 
\label{sec:Variability}
Since the SC and HSC observations with NB816 and i band are separated by $\gtrsim$ 12 years, we check ELGs' two-epoch variation in narrow-band NB816 ($\Delta m_{\rm NB816}\equiv\rm NB816_{HSC} - NB816_{SC}$) and $i$-band ($\Delta m_{i}\equiv i_{\rm HSC} - i'_{\rm SC}$), respectively. 

\begin{figure*}
    {\centering
    \includegraphics[width=1.4\columnwidth]{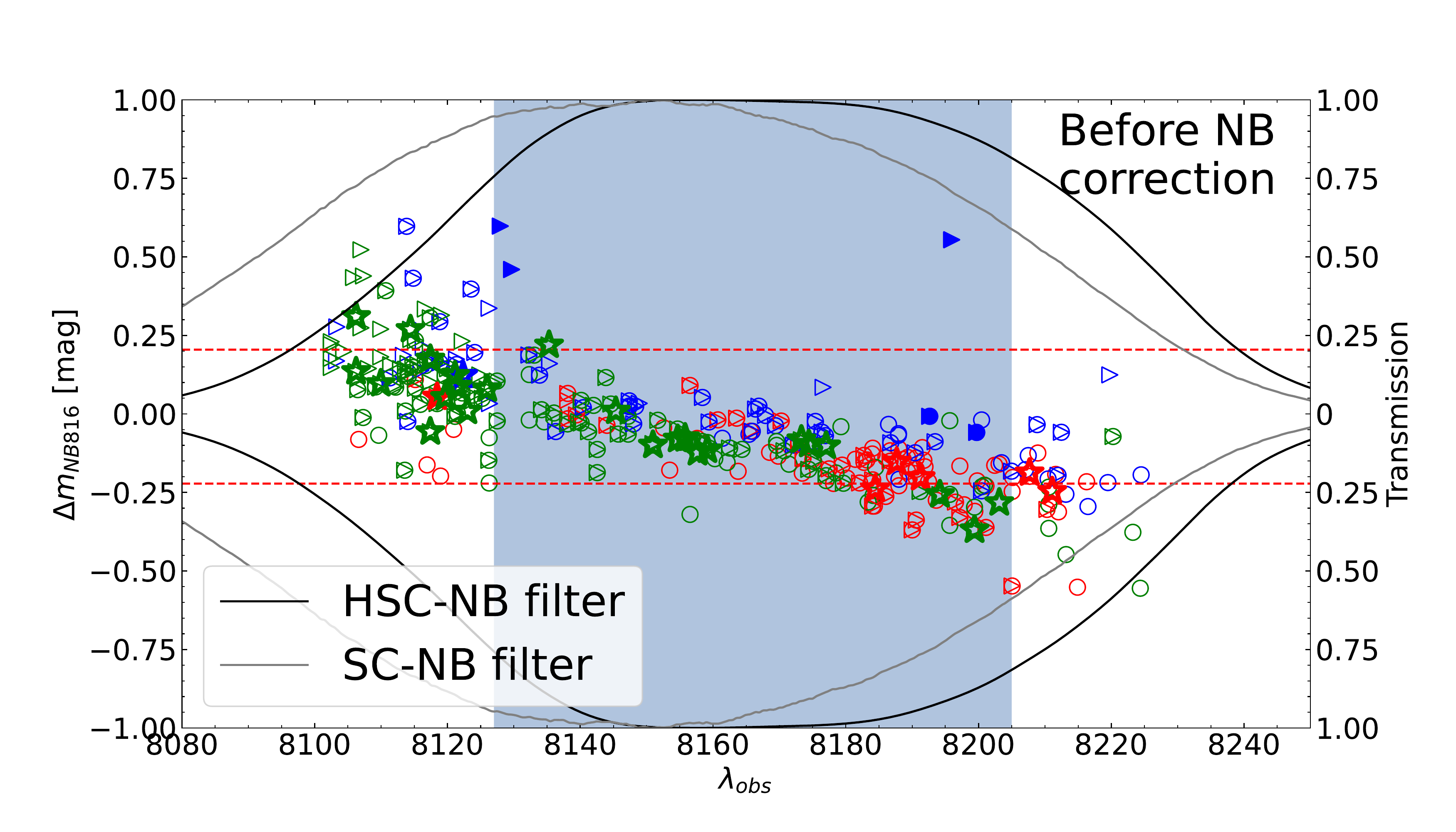}
    \includegraphics[width=1.4\columnwidth]{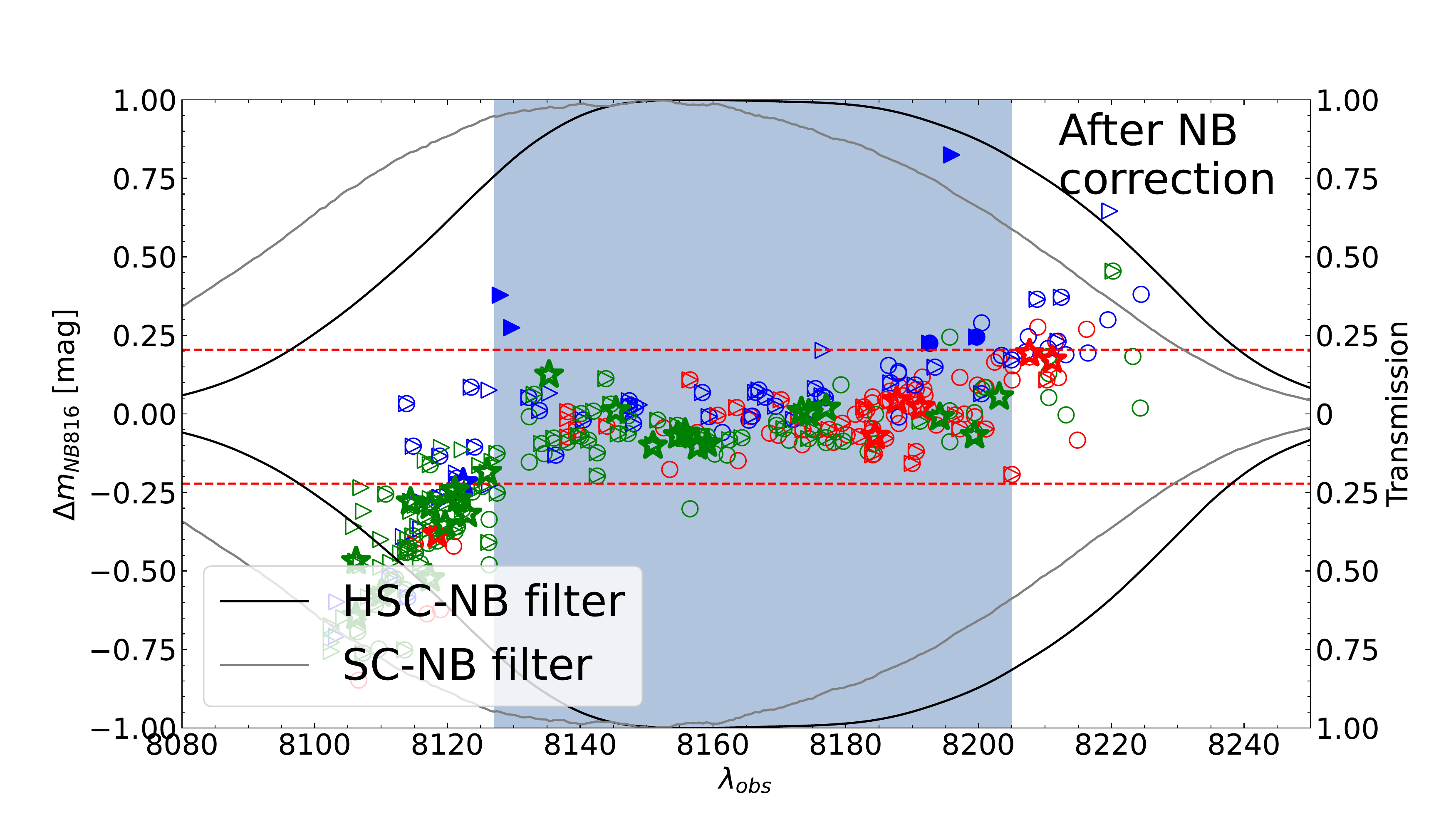}
    \includegraphics[width=1.4\columnwidth]{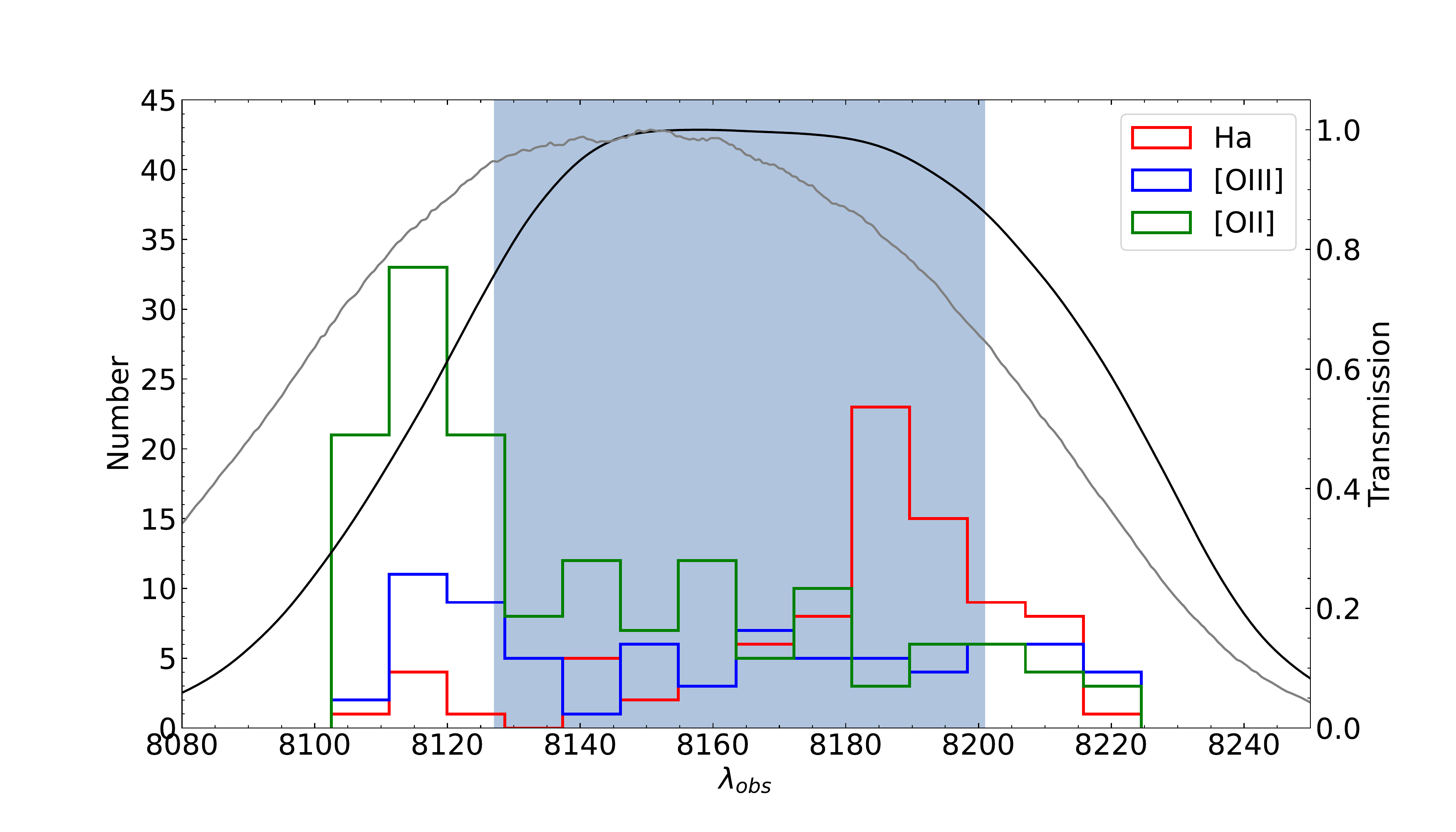}
    \caption{ {\it Top panel}: observed central wavelengths of ELGs' emission-lines $\rm \lambda_{obs}$ versus $\Delta m_{NB816}$ before NB correction. {\it Middle panel}: $\rm \lambda_{obs}$ versus $\Delta m_{NB816}$ after NB correction. {\it Bottom panel}:  Distribution of $\rm \lambda_{obs}$ for H$\alpha$, [O\,{\sc iii}], and [O\,{\sc ii}] emitters. The light blue regions mark the wavelength range of 8127-8205 \AA, which is used to select ELGs as the final ELG sample for the variability test. Red dashed lines show the variability cut of \revise{$\sim$ 0.2} mag. ELGs selected by HSC and SC are marked in open circles and open triangles, respectively. Var-ELGs are marked in solid symbols.
    AGN selected via X-ray, mid-infrared, radio-excess, \revise{and the emission lines diagnostic} methods are marked in stars. Colors shown here are consistent with that in Figure~\ref{fig:z-distirbution}. Transmission curves of the SC-NB816 (gray solid line) and the HSC-NB816 (black solid line) filters are also overlapped in these figures to display the fake variable effect caused by the wing of a narrow-band filter.}
    \label{fig:variable_test}}
\end{figure*}

There is an anti-correlation between $\Delta m_{\rm NB816}$ and the emission lines' central wavelength $\lambda_{obs}$  (see the top panel of Figure~\ref{fig:variable_test}). This is because that the transmission profiles of NB816 filters with HSC and SC are slightly different, therefore the NB flux is sensitive to ELGs with emission-lines centered at the wings of the NB filter (see Figure~\ref{fig:variable_test}). We then correct the NB magnitude by subtracting a term of $\rm 2.5log(1/T_{\lambda_{obs}})$, here $\rm T_{\lambda_{obs}}$ is the transmitted fraction of the filter curve at $\rm \lambda_{obs}$. The corrected NB magnitudes ($\rm NB816_{corrected,\,HSC}\,and\,NB816_{corrected,\,SC}$) are used in the following analysis of optical variability. 

The $\Delta m_{\rm NB816}$ distribution as a function of the emission lines' central wavelength before and after the NB correction are presented in the top panel and middle panel of Figure~\ref{fig:variable_test}, respectively. Obviously, in the central wavelength region overlapped by the two narrow-band filters, the anti-correlation between $\Delta m_{\rm NB816}$ and $\lambda_{obs}$ is corrected. We therefore select the spectroscopically confirmed ELG sample with emission lines located in this wavelength range (from 8127 to 8205 \AA, see the shaded region in Figure~\ref{fig:variable_test}) in the following optical variability test.

\begin{figure}
    {\centering
    \includegraphics[width=\columnwidth]{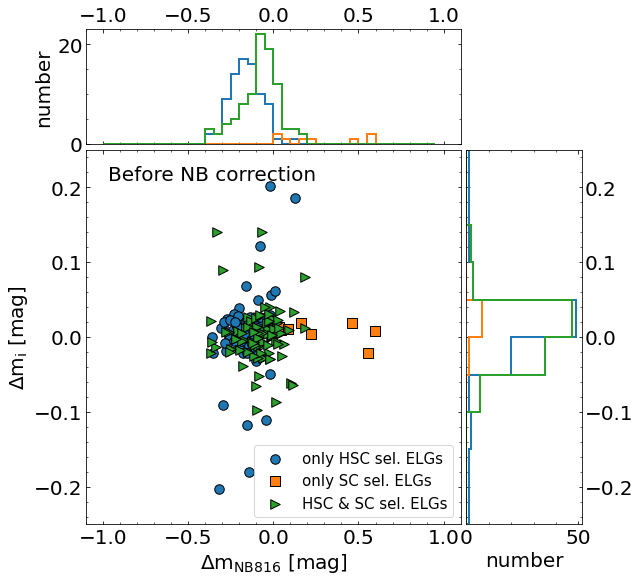}
    \includegraphics[width=\columnwidth]{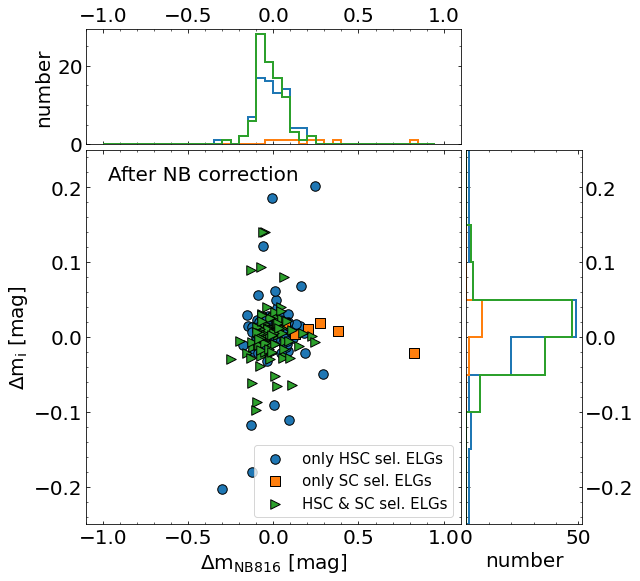}
    \caption{
    Two-epoch magnitude differences in NB and BB of our ELGs before and after the NB correction ({\it upper panel} and {\it lower panel}), respectively.
    The ELGs selected in SC only, in HSC only, and in both SC and HSC are marked as orange squares, blue circles, and green triangles, respectively. The histograms of magnitude differences are also presented.}
    \label{fig:comparison_ELG}}
\end{figure}

We then check the variability of ELGs selected from individual epoch, and both epochs (i.e., selected by both HSC and SC).
\newrevise{Figure}~\ref{fig:comparison_ELG} shows the NB magnitude difference versus the BB magnitude difference for these cases. For all cases, BB magnitude differences $\Delta m_i$ distribute around 0 mag and show no significant systematic bias \newrevise{($|\Delta m_i| = 0.005\pm0.018$).}
However, unlike the distribution of $\Delta m_i$, the distributions of NB magnitude differences $\Delta m_\textrm{NB816}$ are significant before and after the NB correction, especially for those ELGs selected by either HSC or SC. After the NB correction, the distribution of $\Delta m_\textrm{NB816}$ shows no systematic offset \newrevise{($|\Delta m_\textrm{NB816}| = -0.007\pm0.086$)}.


There are \revise{102} out of \revise{181} ELGs with $\Delta m_{\rm NB816} \leq 0$. The ratio of numbers of ELGs brightened and dimmed is approximately \revise{5.6:4.4}. This unequal fraction may be due to the detection image chosen here, which is HSC-NB816 and is deeper than SC-NB816. Sources brightened are more easily detected in HSC-NB816.




\section{Results}
\label{sec:results}

\subsection{Presence of Strong Optical Variability in ELGs}
\begin{figure*}
    {\centering
    \includegraphics[width=1.05\columnwidth]{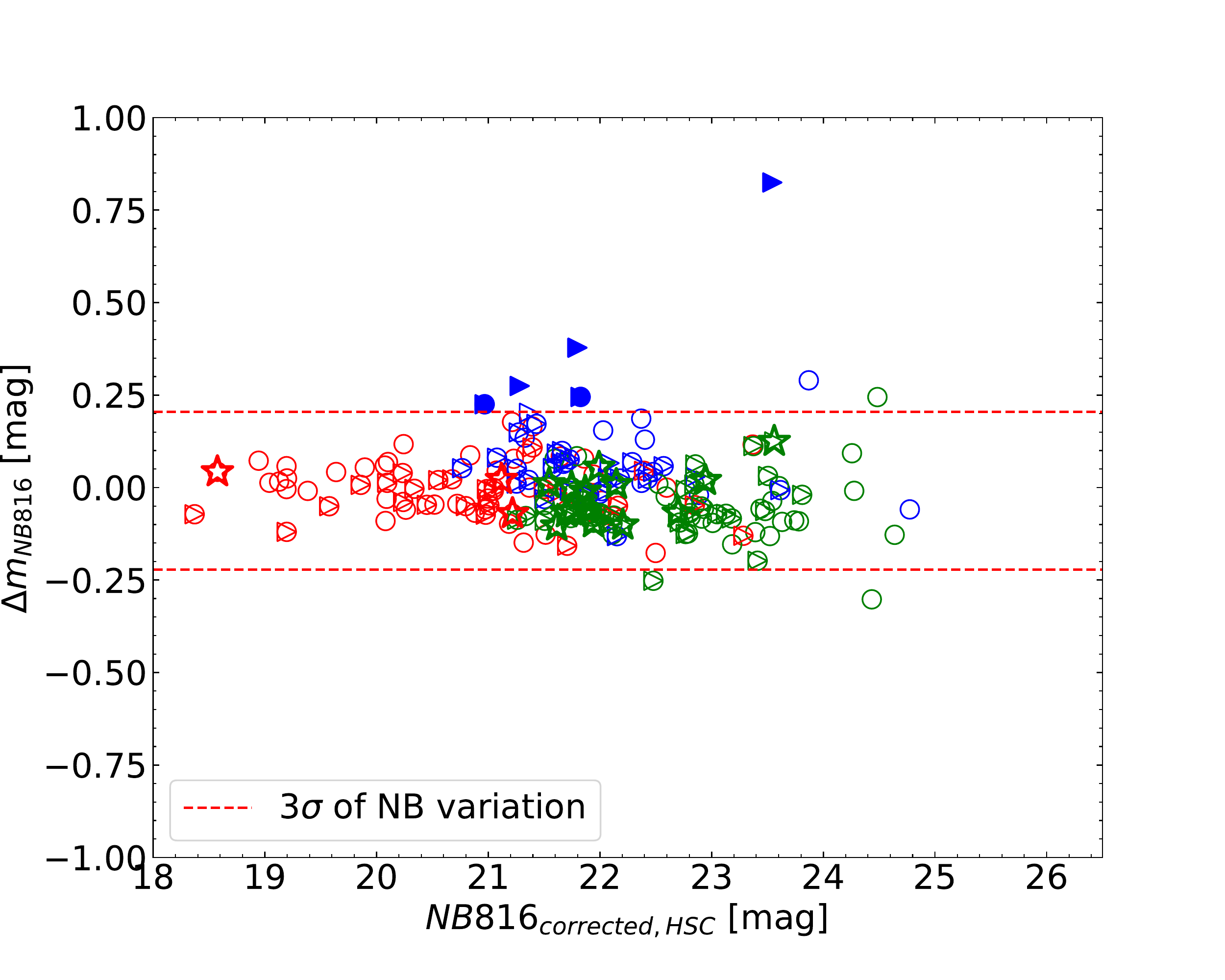}
    \includegraphics[width=1.05\columnwidth]{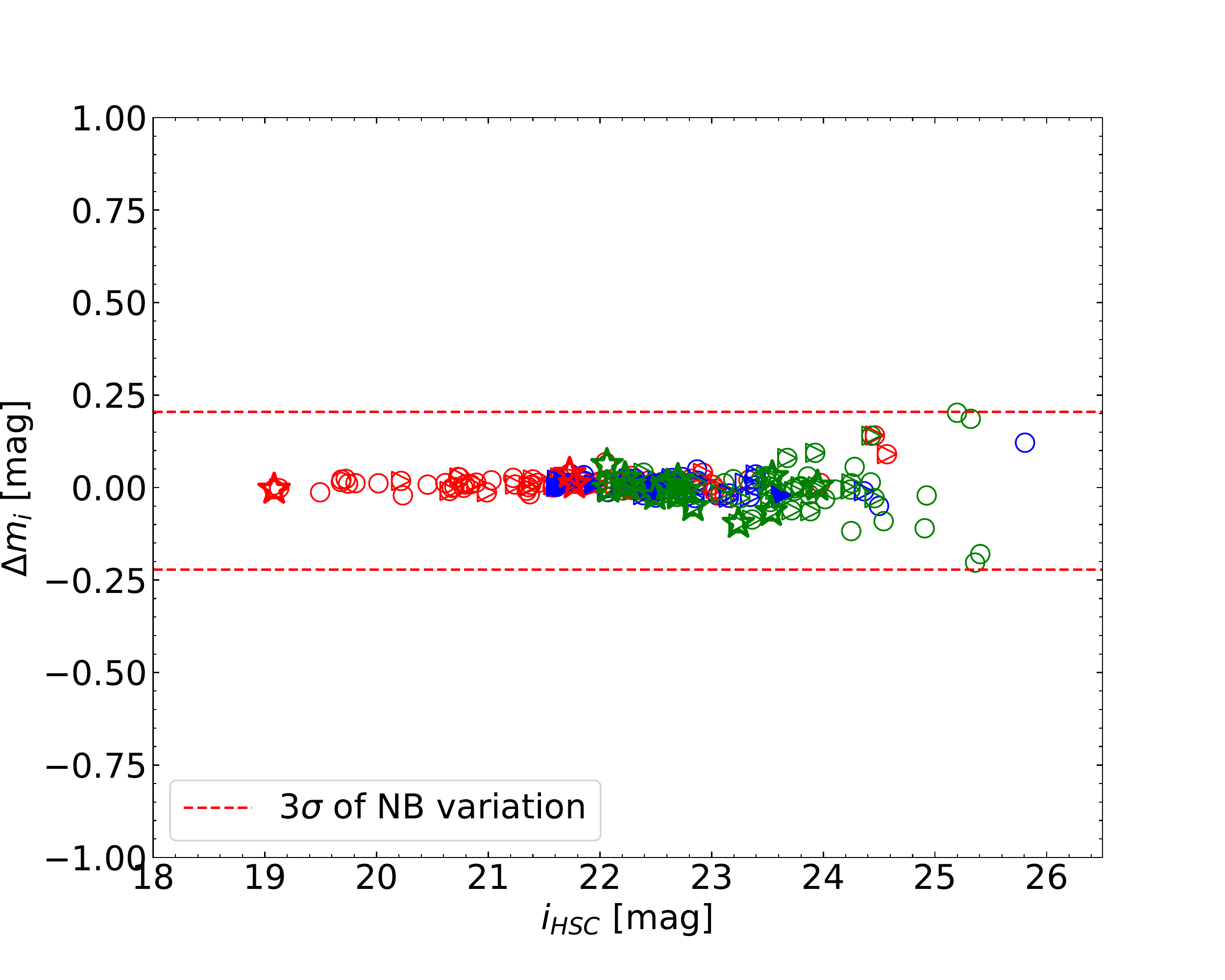}
    \caption{The two-epoch magnitude differences versus magnitudes with HSC of ELGs in the narrow-band ({\it left}) and broad-band ({\it right}), respectively. Both NB-magnitudes and NB-magnitude differences used here are corrected by the NB transmissions. Symbols, lines, and colors here are the same as Figure~\ref{fig:variable_test}.}
    \label{fig:variable_ELG_selection}}
\end{figure*}

The variability test of ELGs in this work are presented in Figure~\ref{fig:variable_ELG_selection}. Obviously, the variability in the narrow-band is more significant than that in the broad-band, which may indicate strong variable emission-lines in these ELGs. However, we should point out that the observing strategies of narrow-band and broad-band are different, which prevent us to check if the variability in narrow-band is caused by the variable emission lines or by the short-term variable continuum. 

\revise{
We then select the strong variable ELGs (hereafter, var-ELGs) in this sample. The selection methods are summarized as below. Firstly, the variability should be significant at S/N $>$ 5. Then the RMS of the two-epoch narrow-band variation of AGN in our ELG sample, selected via four techniques discussed in Sec. \ref{sec:AGN_in_var_ELGs} and \ref{sec:AGN_in_ELGs}, is calculated as $\sigma (\Delta m_{NB816}^\textrm{AGN})$ = 0.067 mag. 
We therefore choose a two-epoch variation cut of $|\Delta m_{NB816} | \geq \revise{0.20}$ mag to select the ELGs with strong variability, corresponding to a search for outliers at $\gtrsim\,3\sigma$ level compared to typical AGN in this ELG sample. }
Only \revise{5} [O\,{\sc iii}] emitters of \revise{181} ELGs meet this criteria, corresponding to a fraction of \revise{2.8\%} (0, \revise{11.9\%}, and \revise{0} for H$\alpha$, [O\,{\sc iii}], and [O\,{\sc ii}] emitters, respectively) of this ELG sample. 



\subsection{Presence of AGN in Variable ELGs}
\label{sec:AGN_in_var_ELGs}
Variability is one of the basic observational features of AGN. To investigate the presence of AGN in this ELG sample, we apply three methods,
including X-ray photometry, mid-infrared (MIR) color, and radio activity. \revise{In addition, we also apply the emission lines diagnostic in a sub-sample of H$\alpha$ emitters}.
X-ray emission indicates the contribution from the accretion of the central massive black hole. MIR color presents the radiation of the dust torus of AGN. Radio activity directly traces the radio jet, which is also powered by the accretion system around the central massive black hole. 
\revise{Emission lines diagnostic (e.g. the BPT diagram, ~\citealt[]{baldwin1981}) can effectively classify the star-forming galaxies and AGN by the hardness of radiation.}
Therefore, these \revise{four} methods would help to effectively select AGN in a galaxy sample.  


Thanks to the abundant multi-band \revise{photometric and spectroscopic survey data} existed in the COSMOS field, we would use these different methods to reveal AGN in our ELG sample, especially the var-ELG sample.  First, we use X-ray data observed with $Chandra$ space telescope to identify AGN in the COSMOS field. Next, we search for AGN by using MIR data taken with $Spitzer$ Space Telescope in this field. We then cross-match our ELG catalog with the COSMOS VLA 3 GHz AGN catalog to select radio-active AGN. \revise{Lastly, we measure the fluxes of H$\alpha$, H$\beta$, [N{\sc ii}] $\lambda$6583, and [O{\sc iii}] $\lambda$5007 lines from available spectra taken from the {\it z}COSMOS DR3 and plot the BPT diagram for identifying AGN from H$\alpha$ ELGs.}

\subsubsection{X-ray Luminosity}
\label{sec:chandra}
 The {\it Chandra COSMOS-Legacy} Survey ~\citep[][]{Civano2016}
is an X-ray survey in the 2.2 deg$^2$ COSMOS field taken with the ACIS instrument on the $Chandra$ space telescope. {\it Chandra COSMOS-Legacy} survey covers the soft (0.5 - 2.0 keV), hard (2.0 - 10.0 keV) and full (0.5 - 10 keV) bands. In the full band, {\it Chandra COSMOS-Legacy} detects 4016 sources with a detection limit of $8.9 \times 10^{-16}\rm \,erg\,cm^{-2}\,s^{-1}$. There are also {\it XMM-Newton} COSMOS survey~\citep[XMM-COSMOS,][]{hasinger_xmm-newton_2007,cappelluti_xmm-newton_2007} in this field with the flux limit of $3.3\times 10^{-15}\rm \,erg\,cm^{-2}\,s^{-1}$ in 2-10 keV band.

Galaxies with X-ray luminosities (full band) of $L_{\rm{FB}} \geq 10^{42}\,\rm{erg\,s^{-1}}$ are commonly regarded as AGN~\citep[e.g., see ][]{Szokoly2004,Finkelstein2009}. 
The detection limit of the {\it Chandra COSMOS-Legacy} survey can be converted to X-ray luminosity limits of $L_{\rm{FB}}=10^{41.2}\,\rm{erg\,s^{-1}}$ for H$\alpha$ emitters ($z$ $\sim$ 0.25), $L_{\rm{FB}}=10^{42.2}\,\rm{erg\,s^{-1}}$ for [O\,{\sc iii}] emitters ($z$ $\sim$ 0.62), and $L_{\rm{FB}}=10^{42.9}\,\rm{erg\,s^{-1}}$ for [O\,{\sc ii}] emitters ($z$ $\sim$ 1.19), implying that all X-ray AGN in H$\alpha$ emitters, and bright X-ray AGN in [O\,{\sc iii}] and [O\,{\sc ii}] emitters should be selected with the current X-ray survey in the COSMOS field.

By checking the footprint of the X-ray surveys, we find that all of \revise{5} var-ELGs are covered by {\it Chandra} and {\it XMM-Newton}, but
none of them show X-ray detection. 
Therefore, in X-ray view, bright AGN are unlikely the origin of var-ELGs.

\begin{figure}
    \centering
    \includegraphics[width=1.08\columnwidth]{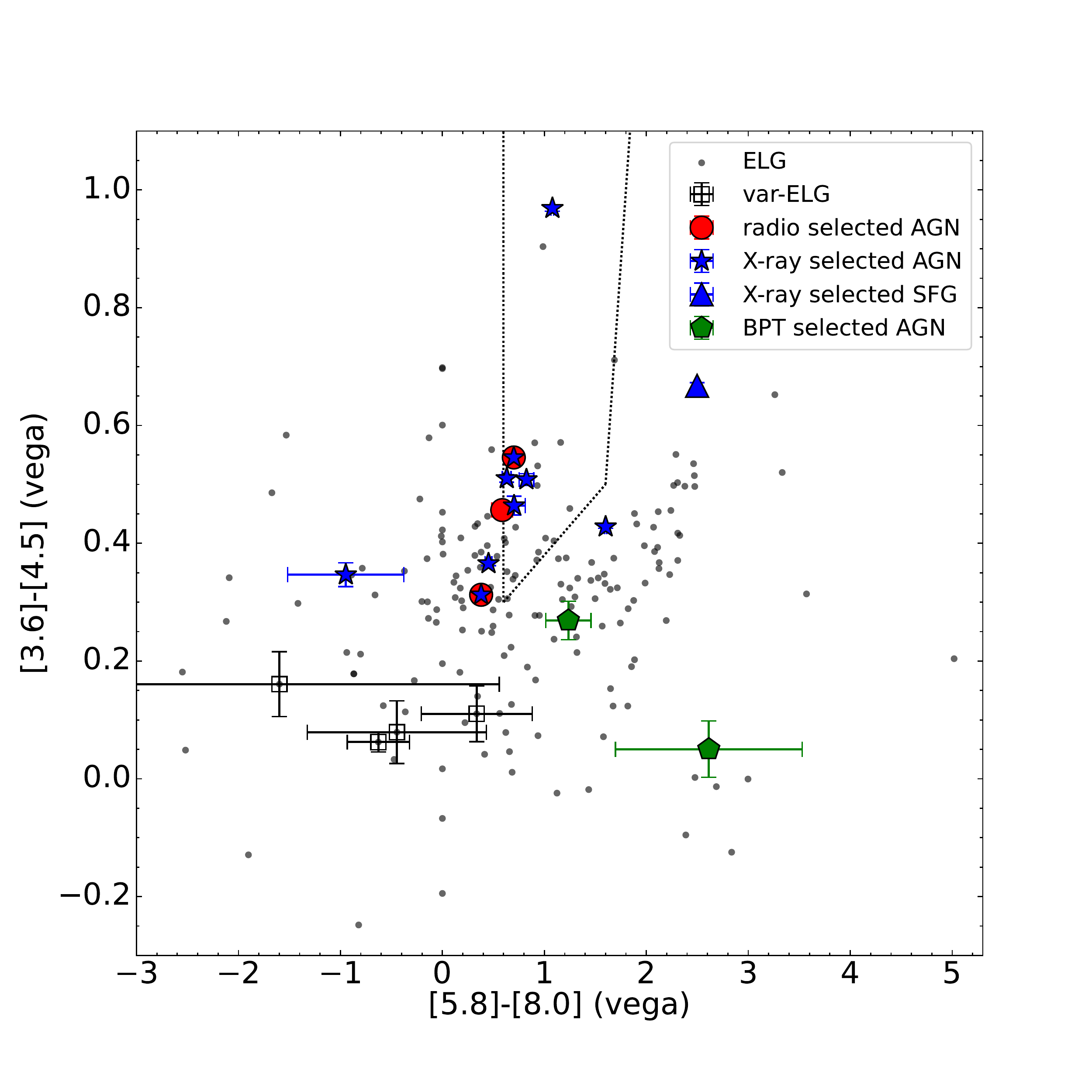}
    \caption{Mid-infrared color-color diagram for AGN selection. ELGs and var-ELGs with IRAC detections in all four bands are marked in gary dots and squares, respectively.
    \revise{4 out of 5} var-ELGs have all four bands detection, and none of them are MIR AGN. X-ray selected AGN and star-forming galaxies are marked in blue stars and blue triangles, respectively. AGN selected by~\citet{Delvecchio2017} via the radio-excess method are marked in red circles. \revise{AGN selected via the BPT diagnostic are marked in green diamonds.} The sources within the dotted line region are regarded as MIR AGN \citep{Stern2005}.}
    \label{fig:MIR_AGN_selection_stern}
\end{figure}
\subsubsection{Mid-Infrared Activity}
The $Spitzer$ COSMOS survey~\citep[S-COSMOS,][]{Sanders2007} is a four channels (3.6, 4.5, 5.8 and 8.0 $\micron$) MIR survey in the COSMOS field taken with the Infrared Array Camera (IRAC) on the $Spitzer$ space telescope.
The observations taken in January 2006 covered the full COSMOS field. 
The 5 $\sigma$ detection limits of the MIR survey are  0.9, 1.67, 11.3, and 14.6\,$\mu \rm Jy$ in 3.6, 4.5, 5.8 and, 8.0 $\micron$ bands, respectively. 

All var-ELGs are covered by the S-COSMOS survey. 
We perform the MIR-AGN selection criteria following~\citet{Stern2005}.
The result is shown in Figure~\ref{fig:MIR_AGN_selection_stern} and Table~\ref{tab:photometry}.
Only \revise{4} var-ELGs have $\gtrsim5\sigma$ detections in all the four MIR bands, while none is selected as MIR-AGN.
The remaining \revise{var-ELG has} non-detection in all MIR bands, implying weak MIR activities.
Therefore, we conclude that no MIR-AGN is found in our variable ELG sample.




\subsubsection{Radio-excess}
~\citet{Delvecchio2017} surveyed 7700 radio sources in the COSMOS field with the Very Large Array (VLA) as the VLA-COSMOS 3 GHz Large Project. They presented an AGN catalog including 1169 sources brighter than the 5 $\sigma$ sensitivities of 2.3 uJy/beam in 3 GHz, along with 1.4 GHz detection~\citep[]{schinnerer_2010}. 
To investigate the presence of AGN, they performed a variety of methods including spectral energy distribution (SED) fitting, X-ray luminosity, MIR activity, and radio-excess. The X-ray and MIR methods they used are similar to what we used in this paper. As for the radio method, they assume that the 1.4 GHz radio excess to the star formation rate indicates the activity of AGN in the host galaxy.

By checking the footprint of the VLA-COSMOS 3 GHz and 1.4 GHz surveys, respectively, we find that all of the var-ELGs are covered. However, none of them is matched with this AGN catalog within a 1.5\arcsec\,radius, indicating no strong radio activity in var-ELGs.





\subsubsection{\revise{{\it Emission Lines Diagnostic}}}
\revise{The spectroscopic redshifts of our ELG sample are collected from the {\it z}COSMOS DR3, 3D-HST, RIMUS, C3R2, and DEIMOS surveys (see \S~\ref{sec:sample} for the detailed references). Due to the different kinds of observational limits (e.g., the wavelength coverage and the exposure depth) of various spectroscopic surveys, only a sub-sample of  ELGs can be probed via the emission line diagnostic systematically. Among these surveys, the {\it z}COSMOS survey has a relatively better spectral quality and a larger coverage for our ELG sample. The {\it z}COSMOS-bright survey releases $\sim$ 10,000 galaxies' spectra in the redshift range of 0.1 < {\it z} < 1.2. A total of 73 spectra from this survey are available for the ELG sample  ($\sim$ 40\% of our ELG sample). Since the wavelength coverage of these spectra is in the range of 5550-9450 \AA, the BPT diagram analysis is limited to the galaxies at redshifts lower than $\sim$ 0.4. Therefore, we only analyze the available spectra of 42 H$\alpha$ emitters here.}

\begin{figure*}
    {\centering
    \includegraphics[width=1\columnwidth]{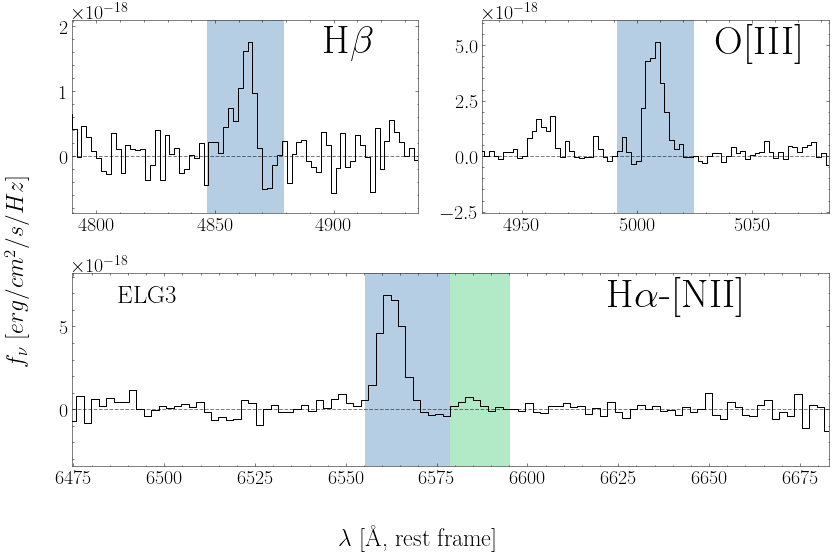}
    \includegraphics[width=1\columnwidth]{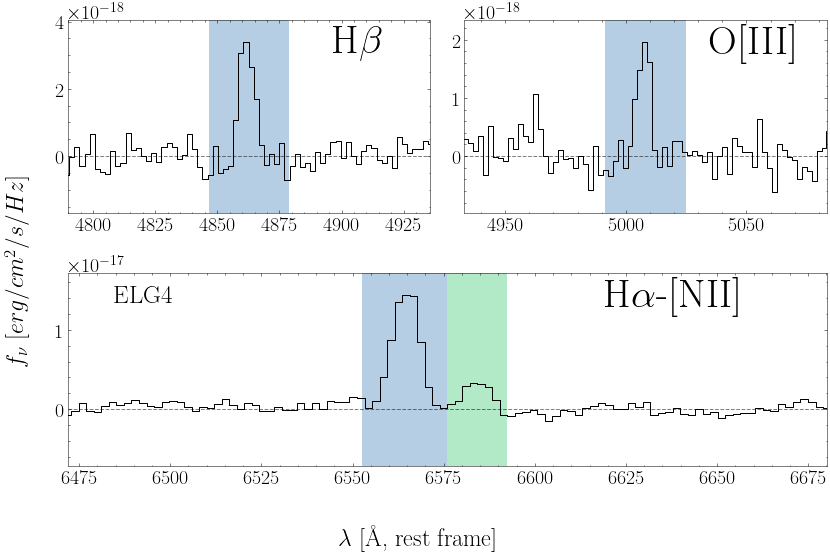}
    \includegraphics[width=1\columnwidth]{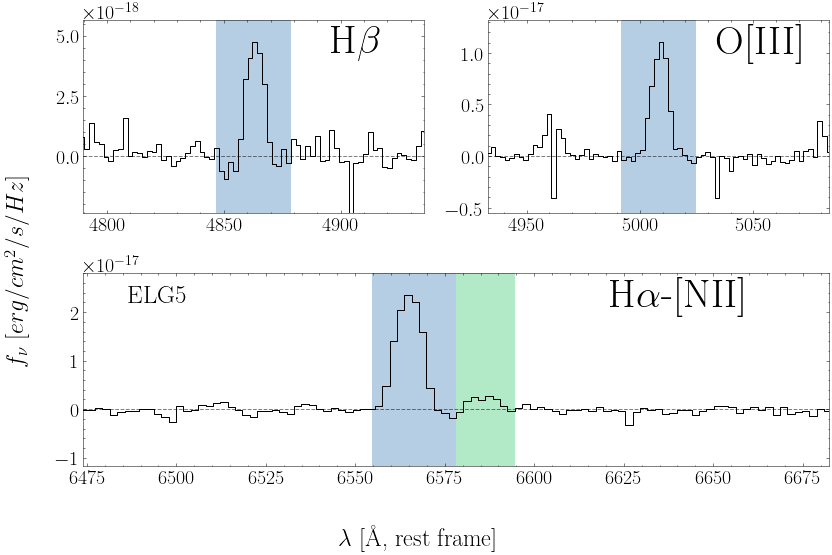}
    \includegraphics[width=1\columnwidth]{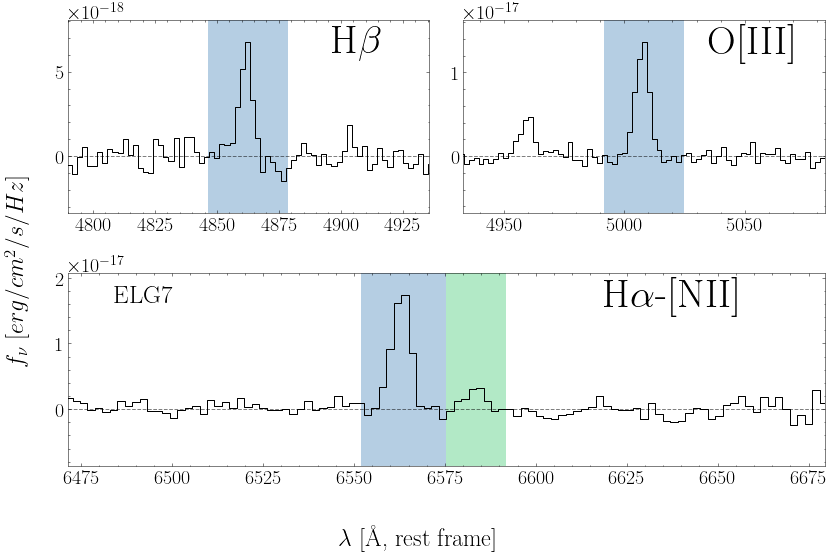}
    \includegraphics[width=1\columnwidth]{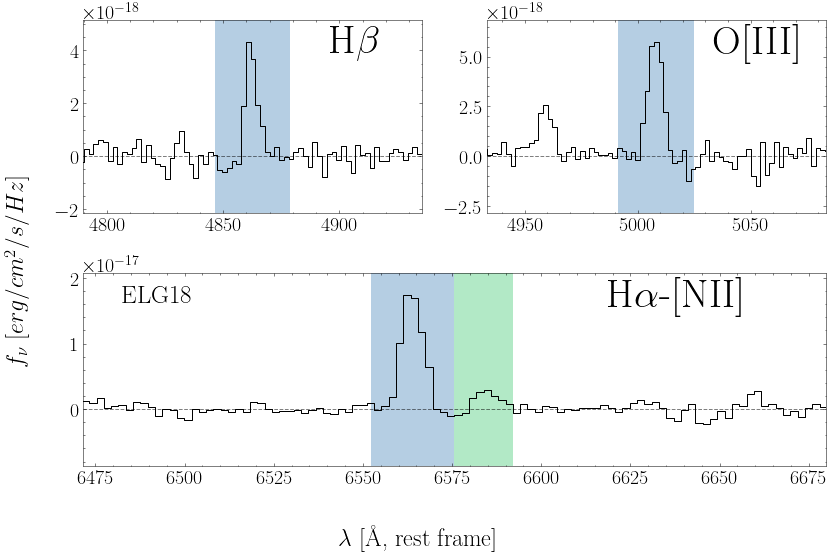}
    \includegraphics[width=1\columnwidth]{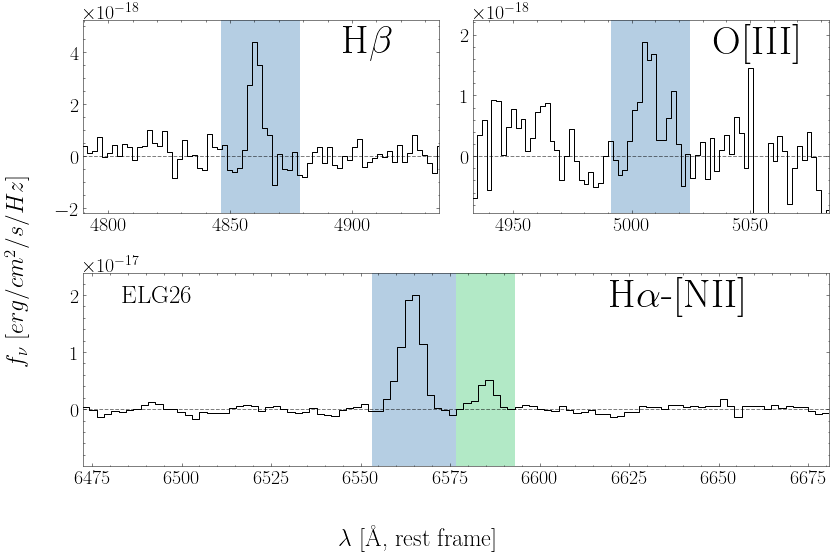}
    \caption{\revise{Examples of measurement of $\rm H\beta, [O III], H\alpha, and [N II]$ emission lines derived from the {\it z}COSMOS spectra. The integrating regions of each emission lines are marked in blue and green shadows.}
    }
    \label{fig:line_measure}}
\end{figure*}

\begin{figure}
    {\centering
    \includegraphics[width=\columnwidth]{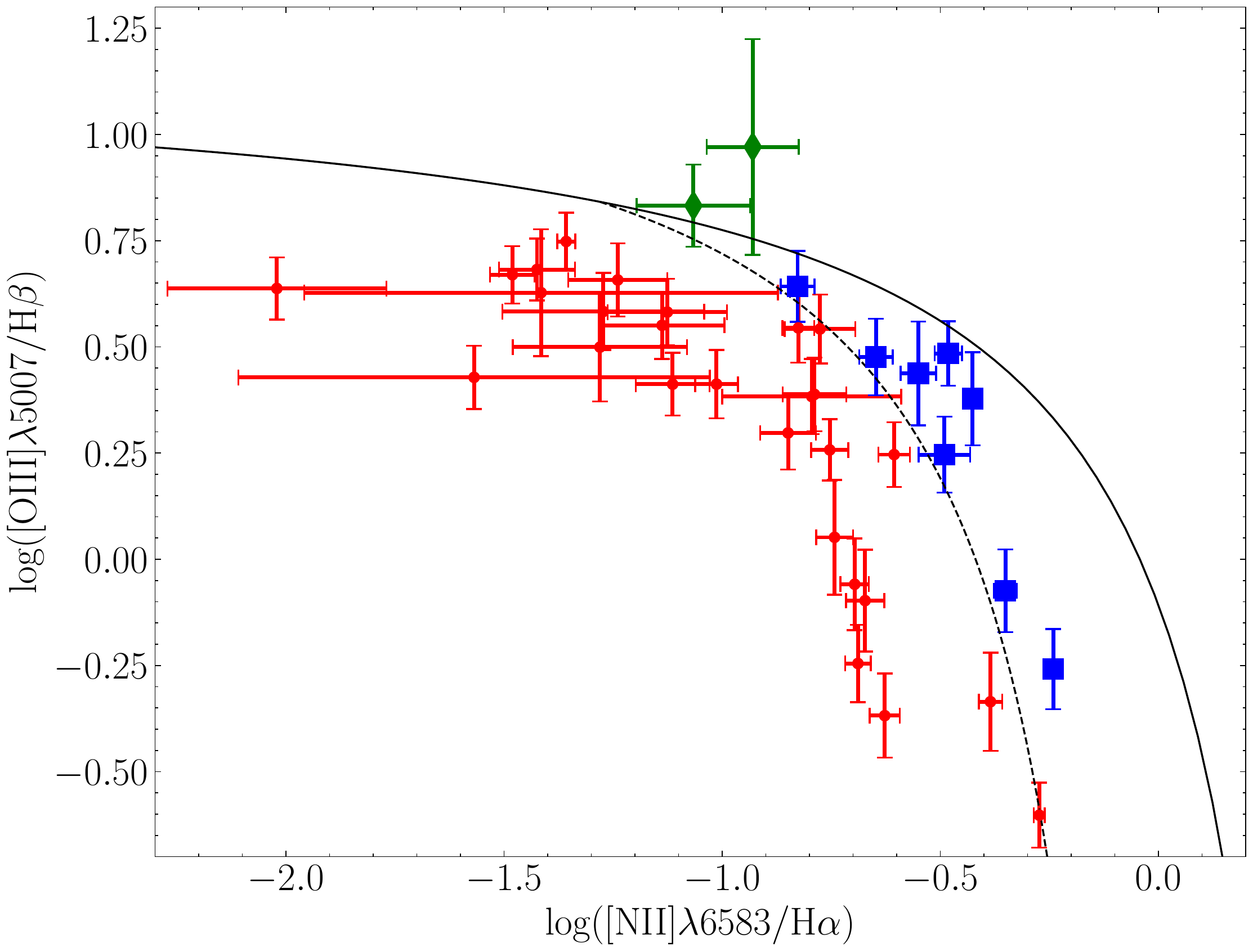}
    \caption{\revise{The BPT diagram of 42 H$\alpha$ emitters covered by the {\it z}COSMOS survey \citep{Lilly2009}. The black solid and dot lines show the traditional BPT diagram separation lines~\citep{baldwin1981, kewley2006}. The red, blue, and green colors are represented the BPT-selected star-forming galaxies, the composite sample, and the BPT-selected AGN, respectively. }  
    }
    \label{fig:BPT}}
\end{figure}

\revise{We firstly measure the fluxes of emission lines including H$\alpha$, H$\beta$, [N{\sc ii}] $\lambda$6583, and [O{\sc iii}] $\lambda$5007 (see Figure~\ref{fig:line_measure} for examples).
The line-flux errors are estimated by computing the RMS in the corresponding continuum region nearby the emission lines.
}
\revise{We then plot the ratios of [N{\sc ii}] $\lambda$6583/H$\alpha$ versus [O{\sc iii}] $\lambda$5007/H$\beta$ (the BPT diagram, ~\citealt{baldwin1981,kewley2006}) in Figure~\ref{fig:BPT} . We identify 2 AGN via the BPT diagram from 42 H$\alpha$ emitters. The final results are given in Table~\ref{tab:ELGs}. } \newrevise{We note that this emission lines diagnostic is not applied to the var-ELG sample because their [N{\sc ii}] $\lambda$6583 and H$\alpha$ lines are not covered by the wavelength range of their spectroscopic surveys. }

\begin{sidewaystable*}
    \centering
    \footnotesize
    \caption{\revise{Physical Parameters of Variable ELGs}}
    \label{tab:photometry}
    \renewcommand\arraystretch{1.2}
    \setlength{\tabcolsep}{0.05cm}{
    \begin{tabular}{lccccccccccccccc} \hline \hline
    \multirow{2}{*}{Var-ID}
    & \multirow{2}{*}{line}
    & \multirow{2}{*}{ra} 
    & \multirow{2}{*}{dec} 
    & corrected HSC 
    & HSC i 
    & corrected SC 
    & SC i  
    & $\rm\Delta m_{NB816}$
    & X-ray 
    & MIR  
    & radio 
    & \multirow{2}{*}{AGN} 
    & \multirow{2}{*}{spec-z}  
    & $\rm SFR_{HSC}^1$ 
    & $\rm SFR_{SC}^1$ 
    \\
      &  &  &  & NB816 (mag) & (mag)  & NB816 (mag)  & (mag) & (mag) & AGN  & AGN  & AGN &  &  & ($M_{\odot}\,\rm yr^{-1}$)& ($M_{\odot}\,\rm yr^{-1}$)
    \\ \hline
    var-ELG1 & [O{\sc iii}] & 150.547364 & 2.804791 & 21.799$\pm$0.020 & 21.949$\pm$0.006 & 21.421$\pm$0.027 & 21.941$\pm$0.034 & 0.378$\pm$0.033 & 0 & 0 & 0 & 0 & 0.623 & 0.14 & 1.71  \\
    var-ELG2 & [O{\sc iii}] & 150.268059 & 2.775813 & 21.282$\pm$0.010 & 21.677$\pm$0.004 & 21.008$\pm$0.022 & 21.659$\pm$0.030 & 0.275$\pm$0.024 & 0 & 0 & 0 & 0 & 0.623 & 1.46 & 3.18  \\
    var-ELG3 & [O{\sc iii}] & 150.067333 & 2.473629 & 23.544$\pm$0.071 & 23.623$\pm$0.023 & 22.720$\pm$0.067 & 23.643$\pm$0.080 & 0.825$\pm$0.097 & 0 & 0 & 0 & 0 & 0.636 & -0.03 & 0.95  \\
    var-ELG4 & [O{\sc iii}] & 150.681233 & 2.763416 & 20.968$\pm$0.010 & 21.610$\pm$0.005 & 20.743$\pm$0.022 & 21.608$\pm$0.031 & 0.225$\pm$0.024 & 0 & 0 & 0 & 0 & 0.636 & 3.46 & 5.54 \\
    var-ELG5 & [O{\sc iii}] & 149.936369 & 1.990318 & 21.827$\pm$0.016 & 22.504$\pm$0.008 & 21.582$\pm$0.036 & 22.511$\pm$0.045 & 0.245$\pm$0.039 & 0 & 0 & 0 & 0 & 0.637 & 1.65 & 2.72  \\ 
    \hline 
    \end{tabular}
    }
    \begin{tablenotes}
    \footnotesize
    \item Column 1: ID of variable ELGs in this paper; Column 2: emission-line type; Column 3-4: coordinates (equatorial); Column 5 and 7: corrected NB816 magnitude with HSC and SC as describe in section.~\ref{sec:Variability}; Column 6 and 8: broadband magnitude with HSC and SC; Column 9: NB816-magnitude differences between NB816 observations with HSC and SC. Column 10-12: X-ray, MIR and radio selected AGN. 0 means non-detection; Column 13: total number of AGN via above three methods; Column 14: spectroscopic redshift; Column 15-16: two-epoch SFRs of variable ELGs calculated by SFR-emission-line indicators. 
    $^1$ SFR([O{\sc iii}]) following~\citet{Zhuang2019}.
    \end{tablenotes}
\end{sidewaystable*}

\subsubsection{Final Classification}
As shown in Table~\ref{tab:ELGs} and Table~\ref{tab:photometry}, none of the var-ELGs is identified as AGN via X-ray luminosity, MIR activity, or radio-excess methods. This result indicates that strong variability in ELGs are not dominated by AGN. \revise{However, we should note that the faint AGN would be missed in our var-ELG sample with these three methods when considering their detection limits.}

\subsection{Morphology}
\label{sec:morphology}
\begin{figure*}
\centering
\includegraphics[width=0.7\linewidth]{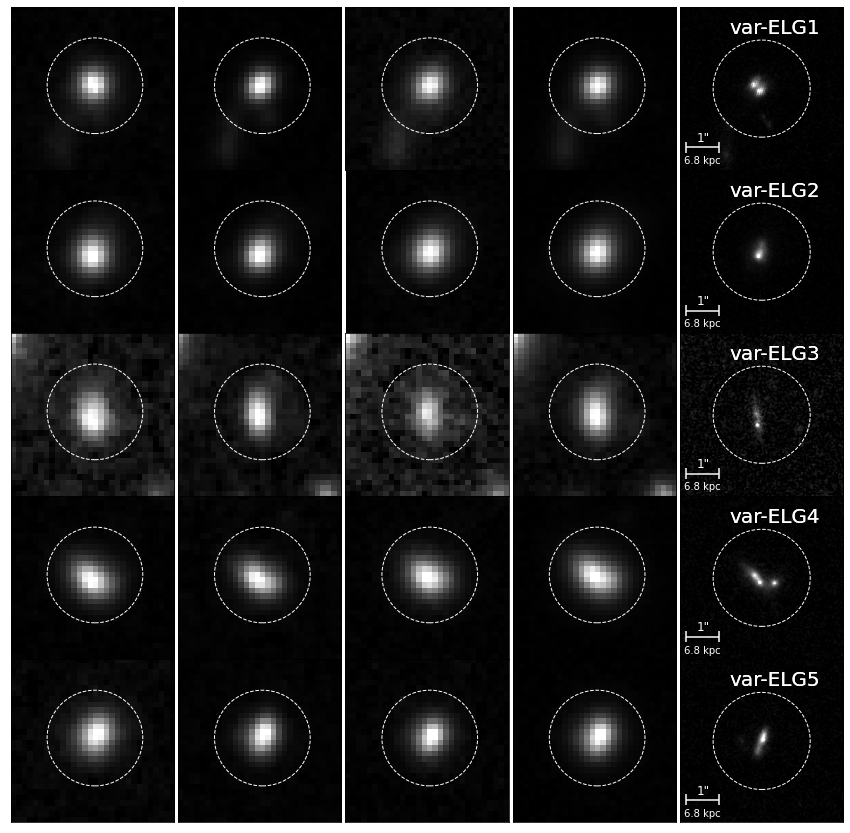}
\caption{Image cutouts of \revise{5} var-ELGs. All stamps are in a size of 5\arcsec $\times$ 5\arcsec. Column 1-2:   NB816 and $i'$-band with SC; Column 3-4: NB816 and i-band with HSC; Column 5:  HST/F814W band. The circles with white dashed lines have 3\arcsec\ diameters.}

\label{fig:morpholigy1} 
\end{figure*}



We use HST/ACS F814W imaging data to check the morphology of var-ELGs. Their morphology in NB816 and $i'/i$ with SC and HSC are also checked. Figure~\ref{fig:morpholigy1} shows the image cutouts of var-ELGs, including the NB816 and $i(i')$ images with HSC(SC), and HST/F814W images. 

With HST, \revise{all} of the var-ELGs show the structures of mergers (var-ELG \revise{1, 3, 4, \& 5}) or post-mergers (var-ELG \revise{2}), indicating star-forming activities. \revise{Sorted by their identifications, the separations of these galaxy pairs are $\lesssim$ 0.27\arcsec - 0.45\arcsec, corresponding to physical sizes of $\lesssim$ 1.8 - 3.2 kpc. }

\section{Discussion}
\label{sec:discussion}
\subsection{Fractions of AGN}
\label{sec:AGN_in_ELGs}
We have applied three methods, including X-ray luminosity, MIR activity, and radio excess to identify AGN among var-ELGs in \S~\ref{sec:results}. \revise{We also apply the BPT diagnostic to a sub-sample of H$\alpha$ emitters.} Combining these \revise{four} methods, we find \revise{17 AGN} in the total \revise{181} ELGs (see Table~\ref{tab:ELGs}, Figure~\ref{fig:MIR_AGN_selection_stern}, and Figure~\ref{fig:BPT} for details).
The total AGN fractions are $\sim$ \revise{4.4}\%, 0\%, and \revise{19.7}\% for H$\alpha$ emitters at $z$ $\sim$ 0.24, [O\,{\sc iii}] emitters at 0.63, and [O\,{\sc ii}] emitters at 1.19 , respectively. These AGN are marked with stars in Figure~\ref{fig:variable_test} and Figure~\ref{fig:variable_ELG_selection}.

\begin{figure}
    \centering
    \includegraphics[width=1.08\columnwidth]{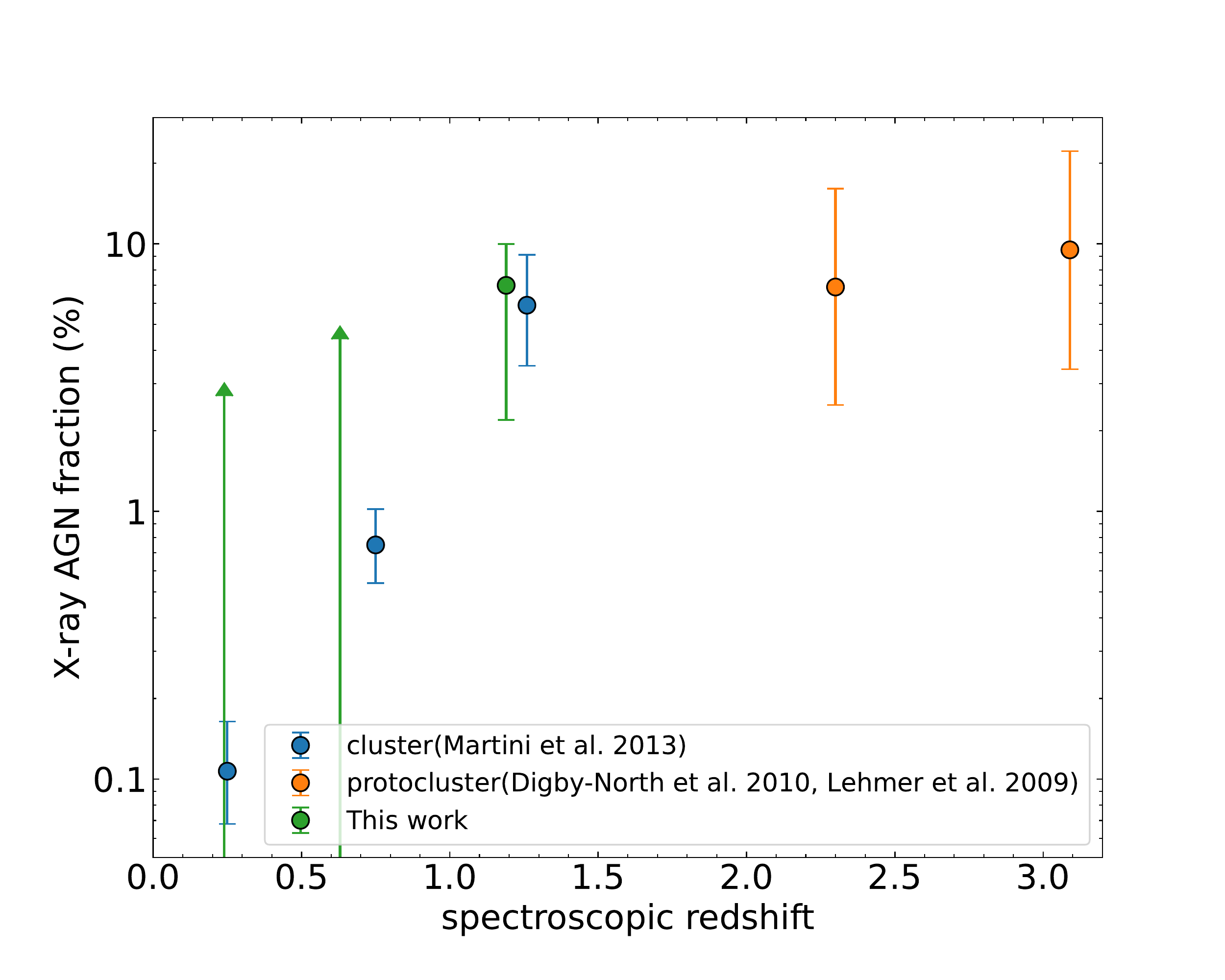}
    \caption{Evolution of the luminous X-ray AGN fraction from $z$ = 0 to $\sim$3. The fraction at $z$ = 0.25 are computed by~\citet{Martini2013} from combining sample of~\citet{Martini2009} and \citet{Haines2012} .The data point at $z$ = 0.75 is computed by~\citet{Martini2009}. The data point at $z$ = 1.26 is calculated by~\citet{Martini2013}. Points at $z$ = 2.3 and 3.09 are AGN fractions in protocluster computed by~\citet{Digby-North2010} and~\citet{Lehmer2009}.  Green arrows and circle indicate the upper limits and value of X-ray AGN fraction of $<$2.7\%, $<$\revise{4.4}\%, and \revise{7.0$^{+4.8}_{-3.0}$}\% in our ELGs at $z$ $\sim$ 0.24, 0.63, and 1.19, respectively.}
    \label{fig:AGN_fraction}
\end{figure}

 To check if our spectroscopically confirmed ELG sample is biased on AGN selection, we compare the X-ray AGN fraction in our sample to that in previous works. For consistence, the criterion for X-ray AGN is set to full-band (0.5-10keV) X-ray luminosity $L_{FB}\gtrsim10^{43}\, erg\, s^{-1}$ \citep[][, see Table~\ref{tab:ELGs}]{Martini2013}.
 The evolution of the X-ray AGN fraction from $z$ = 0.25 to 3.09 is presented in Figure~\ref{fig:AGN_fraction}. The AGN fractions at $z$ = 0.25, 0.75 and 1.25 were computed by \citet{Martini2013} and other previous works~\citep[e.g.][]{Martini2009, Haines2012, Digby-North2010, Lehmer2009} from galaxy clusters. The AGN fraction at $z$ = 2.30 and 3.09 were measured in galaxy protoclusters~\citep[also see~\citealt{Martini2013}]{Digby-North2010, Lehmer2009}.
 In our ELG sample, we find 0 X-ray AGN for either H$\alpha$ emitters or [O\,{\sc iii}] emitters, and 5 X-ray AGN for [O\,{\sc ii}] emitters. We then estimate the AGN fractions of $<$2.7\%,  $<$\revise{4.4}\%, and \revise{7.0$^{+4.8}_{-3.0}$}\% for ELGs at $z$ $\sim$ 0.24, 0.63, and 1.19, respectively. Errors and 1-$\sigma$ upper-limits of fractions are calculated following~\citet{gehrels1986}. The X-ray AGN fractions in our sample are consistent with those in previous works. 
 
\revise{As given in \S~\ref{sec:results}, the fraction of variable sources in ELGs ($\sim$ 2.8\%) is several times lower than the fraction of AGN in ELGs. Furthermore, none of the var-ELGs has been identified as AGN through X-ray luminosity, MIR activity, or radio excess. This infers that bright AGN are unlikely the main origin of the variability of our var-ELG sample. However, due to the detection limits of various methods, we can not rule out the contribution from faint AGN in these var-ELGs. } 


 
\subsection{Stellar-driven Variability?}
\label{sec:SNrate}
\revise{After excluding bright AGN as the origin of var-ELGs, we check the possibility of stellar-driven variable events, such as SNe and TDEs. The variability caused by the extra flux of such event, as well as its event rate, is discussed here. Note that since the observations in narrow-band and broad-band of this sample lack coherence (Table \ref{tab:observing_date}), below we assume that the variability in narrow-band is caused by the short-term variable continuum caught in either epoch of the narrow-band observation. }

It is critical that whether the extra fluxes yielded by SNe and TDEs are significant enough to provide the $\gtrsim$ \revise{0.20} mag variability for ELGs at z = 0.63. For type Ia SNe (SNe Ia) with a typical \newrevise{peak} absolute magnitudes of $M_B \sim$ -19.3 mag~\citep[]{branch_type_1998}, the extra flux it can contribute is $\rm \sim 4 \times 10^{-15}\, erg\,cm^{-2}\,s^{-1}$ at {\it z} = 0.63 by assuming a flat spectrum. The specific values of magnitude differences caused by SNe explosions are then estimated with the apparent magnitudes of the host galaxies. For instance, in galaxies with apparent magnitudes of \revise{21 (or 23.5) mag}, the SNe having a \newrevise{peak} $M_B \sim$ -19.3 mag will result in variations of $\sim$ \revise{-0.10 (or -0.72)} mag at {\it z} $\sim$ 0.63. While by considering the absolute magnitudes of type II SNe (SNe II), which ranges from -17 up to -22 mag \newrevise{at peak}~\citep[]{kiewe_caltech_2012,taddia_carnegie_2013,reynolds_sn_2020,tartaglia2020}, type II SNe can produce the magnitude variation up to \revise{-0.82 (or -2.72) } mag for galaxies with magnitudes of 21 (or 23.5) mag at {\it z} $\sim$ 0.63.


\revise{With above estimations, the SN Ia hypothesis can only explain the variability of var-ELG3, which changed from 23.5 to 22.7 in the two-epoch narrow-band observations required an event with the absolute magnitude of $\gtrsim$ -19.3 mag. The remaining 4 var-ELGs had the narrow-band magnitudes in the range of 20.7 to 21.8 with two-epoch narrow-band magnitude differences of 0.22-0.38, requiring events with absolute magnitudes of $\gtrsim$ -19.5 - -20.3 mag. These values of the magnitudes and two-epoch magnitude differences are in agreement with the hypothesis of SNe II driven variability. \newrevise{However, the required transients would be extreme SNe II events at the bright end which could be quite rare.} Overall, above estimations imply that luminous SNe II would bring in the strong variability of our var-ELGs, while we could not rule out the contribution from faint AGN or other luminous transient events, e.g., the ultraluminous fast blue optical transient (FBOT) reported recently by \citet{jiang2022}. 
}



Event rates of these stellar-driven transients are also very helpful for us to understand the origin of variable ELGs. We check the event rate of two types of SNe, type Ia SNe and core-collapse SNe, respectively. 
Since our sample is a sub-kind of galaxies selected with their strong emission lines at the corresponding redshifts, we estimate the number of SNe per galaxy per year ($R_{SN}$), defined as:
\begin{equation}
    R_{SN}=\rm SNR\times M_{*}.
\end{equation}
Here the SNR is in the unit of \revise{$\rm yr^{-1}\,M_\sun^{-1}$} and  the \revise{$\rm M_{*}$ is the stellar masses of galaxies in the unit of $\rm M_\sun$}. 
According to the SN rate-size relation for SNe Ia and SNe II/Ibc galaxies~\citep{li2011}, the SN rate per unit mass could be described as:
\begin{eqnarray}
log(\frac{\rm SNR(\textrm{SNe Ia})}{10^{12}}) & = & -0.50\times log(\frac{M_{*}}{10^{10}})-0.65,\\
 log(\frac{\rm SNR(\textrm{SNe II/Ibc})}{10^{12}}) & = & -0.55\times log(\frac{M_{*}}{10^{10}})-0.073.
\end{eqnarray}

\revise{We adopt a median stellar mass ($10^{9.35}\ \rm M_\odot$) of our ELGs as the typical mass of ELG. Thus we can derive the SNe event rate ($R_{SN}$) of our ELG sample as: 
$0.46\times10^{-12} \rm M_\odot^{-1}\,yr^{-1}$ for SNe Ia, and
$1.88\times10^{-12} \rm M_\odot^{-1}\,yr^{-1}$ for SNe II or SNe Ibc, corresponding to the total event rate (SNe Ia + SNe II + SNe Ibc) of $\sim$ 0.01 $\rm yr^{-1}$. With the estimations above, there are $\sim$ 1.8 SNe expected in the ELG sample (in total 181 ELGs) per year. This estimated event rate of SNe is $\sim$ 2.8 times lower than the fraction of var-ELGs in the ELG sample. However, the SN rates used here are measured in the local university~\citep{li2011}, while our ELGs have redshifts of $\gtrsim$ 0.24. Furthermore, the SN rate may be higher in the our var-ELG sample, because of their strong interactions between galaxies (see \S~\ref{sec:merger} for details). With the above discussion, SNe may play an important role in contributing to the variability of var-ELGs.} 

TDE would turn the disrupted star into the extra fuel to the active/inactive accretion disk possessed by the central supermassive black hole of the host galaxy. According to \citep[]{gezari_luminous_2009}, the TDE flare could be as energetic as 
$\sim\ 1.6\times10^{43}\ \rm erg\,cm^{-2}\,s^{-1}$ at $z\sim$ 0.19. The corresponding extra fluxes could be $\sim\ 1\times10^{-14}\, erg\,cm^{-2}\,s^{-1}$ at {\it z} = 0.63. With the similar estimation above, the magnitude variations caused by TDEs could also introduce the magnitude differences of our ELG sample.  However, compared to event rates of SNe, the TDE rate is much lower, expected a few times $10^{-5}$ per galaxy per year (see \citealt{gezari2021}, corresponding to $\sim\,2\times10^{-3}$ var-ELGs per year for our ELGs sample). Overall, \revise{SNe are more likely to be the origin of strong optical variability than TDEs in our ELG sample.}



\subsection{\revise{The Merging Features of Var-ELGs}}
\label{sec:merger}
\revise{As described in \S~\ref{sec:morphology}, all the var-ELGs show merging or post-merged structures. Their angle separations of galaxy pairs are less than 0.5\arcsec, corresponding to projected separations of $\lesssim$ 3.4 kpc. These kind structures may imply an underlying correlation between the very close galaxy-galaxy interaction and variability of ELGs. }

\revise{Several works had pointed out that the interaction between galaxies can significantly enhance the AGN and star-forming activity~\citep[e.g.,][]{kennicutt1984,alonso2007}. As the results given in \S~\ref{sec:AGN_in_var_ELGs} and \S~\ref{sec:AGN_in_ELGs}, none of the bright AGN found in the var-ELG sample, indicating that the bright AGN are not the main origin of the strong variability of our var-ELGs. 
However, we could not rule out the contribution from faint AGN.}

\revise{Observations of SDSS galaxy pairs had shown that the smaller the projected separation, the stronger the enhancement of star formation~\citep{li2008,patton2013}, especially in less massive galaxies~\citep{li2008}. That is consistent with the result given in \S~\ref{sec:morphology}. As all of the var-ELGs have small projected separations ($\lesssim$ 3.4 kpc) and low masses ($\lesssim 10^{10} \rm\ M_\sun$), they are more likely to have higher enhancements in star formation. This is in agreement with the picture of the SN-driven variability.}


\section{Conclusion}
\label{sec:conclusion}

We use the two-epoch narrow-band imaging, separated by $\gtrsim$ 12 years, to check the strong optical variability of a sample of \revise{181} spectroscopically confirmed ELGs in the COSMOS field. 
The two-epoch narrow-band imaging was observed with HSC and SC instruments of {\it Subaru} telescope. This sample includes 68 H$\alpha$, \revise{42} [O\,{\sc iii}] and \revise{71} [O\,{\sc ii}] emitters at redshift $z\,\sim$ 0.24, 0.63 and 1.19, respectively. Only \revise{5 [O\,{\sc iii}] emitters} show significant variability in narrow-band ($|\Delta m_\textrm{NB}| \geq 3\,\sigma_{\Delta m_\textrm{NB,AGN}} = \revise{0.20}\, \textrm{mag}$). The fractions of var-ELGs are \revise{2.8}\% of the whole ELG sample.

We probe the existence of AGN in this ELG sample via X-ray luminosity, mid-infrared activities, and radio-excess. We also apply the emission-line diagnostic in a sub-sample of H$\alpha$ emitters. We find no \revise{bright} AGN in the var-ELG sample, indicating that the strong optical variability of our ELGs is not dominated by \revise{bright} AGN. \revise{However, we cannot rule out the contribution from faint AGN.} 

\revise{We discuss the possibility of SNe contributing to the strong variability of var-ELGs. This strong variability could be qualitatively explained by the luminous SNe. }

\revise{With HST images, we find that all of the var-ELGs have merging or post-merged features with projected separations of $\lesssim$ 3.4 kpc. The strong interaction between galaxies may indicate enhanced star formation activity.}

\vspace{15mm}
\section{Acknowledgments}
We would like to thank the referee for very helpful comments and suggestions.
We would like to thank Junxiang Wang, Linhua Jiang, and Yicheng Guo for very useful discussion. Z.Y.Z. acknowledges support by the National Science Foundation of China (12022303), the China-Chile Joint Research Fund (CCJRF No. 1906), and the CAS Pioneer Hundred Talents Program. J.W. is supported by NSFC (grants U1931131). F.T.Y. acknowledges the support from the Funds for Key Programs of Shanghai Astronomical Observatory and the Natural Science Foundation of Shanghai (Project Number: 21ZR1474300).  R.P.T. thanks the CAS President's International Fellowship Initiative (PIFI) (Grant No. E085201009) for supporting this work. 

This paper is based on data collected at the Subaru Telescope and retrieved from the HSC data archive system, which is operated by the Subaru Telescope and Astronomy Data Center (ADC) at NAOJ. Data analysis was in part carried out with the cooperation of Center for Computational Astrophysics (CfCA), NAOJ. We are honored and grateful for the opportunity of observing the Universe from Maunakea, which has the cultural, historical and natural significance in Hawaii. The Hyper Suprime-Cam (HSC) collaboration includes the astronomical communities of Japan and Taiwan, and Princeton University. The HSC instrumentation and software were developed by the National Astronomical Observatory of Japan (NAOJ), the Kavli Institute for the Physics and Mathematics of the Universe (Kavli IPMU), the University of Tokyo, the High Energy Accelerator Research Organization (KEK), the Academia Sinica Institute for Astronomy and Astrophysics in Taiwan (ASIAA), and Princeton University. Funding was contributed by the FIRST program from the Japanese Cabinet Office, the Ministry of Education, Culture, Sports, Science and Technology (MEXT), the Japan Society for the Promotion of Science (JSPS), Japan Science and Technology Agency (JST), the Toray Science Foundation, NAOJ, Kavli IPMU, KEK, ASIAA, and Princeton University. This research uses data obtained through the Telescope Access Program (TAP), which has been funded by the TAP member institutes. We acknowledge the use of the Lijiang 2.4 m telescope and its technical operation and maintenance team. This research uses data obtained through the Lijiang 2.4 m Telescope, which is funded by the Chinese Academy of Sciences, the People's Government of Yunnan Province and the National Natural Science Foundation of China.

This research has made use of the NASA/IPAC Infrared Science Archive, which is funded by the National Aeronautics and Space Administration and operated by the California Institute of Technology. The archival data we use including Chandra-COSMOS Bright Source Catalog~\citep{cosmos_chandra2009}, S-COSMOS IRAC 4-channel Photometry Catalog~\citep{cosmos_irac2007}, and COSMOS 3GHz AGN Catalog~\citep{cosmos_vla2017}.

This paper makes use of software developed for Vera C. Rubin Observatory. We thank the Rubin Observatory for making their code available as free software at http://pipelines.lsst.io/.



%

\vspace{5mm}
\facilities{COSMOS, Subaru(HSC and SC), IRSA, HST(ACS), Chandra, Spitzer, VLA}


\software{astropy\citep{collaboration2013,astropy2018},
          Source Extractor \citep{Bertin1996},
          SWarp \citep{bertin2010},
          TOPCAT \citep{Taylor2005}
          }



\bibliography{reference}{}
\bibliographystyle{aasjournal}

\appendix
\section{Extra Tables}

\begin{table}
\centering
\caption{SExtractor Parameters Used in the Duel-mode}
\label{tab:sext_para}
\setlength{\tabcolsep}{0.5mm}{
\begin{tabular}{llr} 
\hline \hline
parameter         &  &value                 \\ \hline
--------------------&  Extraction &--------------------\\
DETECT\_TYPE      &  &CCD                   \\
DETECT\_MINAREA   &  &5                     \\
DETECT\_MAXAREA   &  &100000                \\
DETECT\_THRESH    &  &0.6 (in sigma)                   \\
ANALYSIS\_THRESH  &  &0.6 (in sigma)                   \\
FILTER            &  &Y                     \\
FILTER\_NAME      &  &gauss\_2.5\_5x5.conv  \\
DEBLEND\_NTHRESH  &  &64                    \\
DEBLEND\_MINCONT  &  &0.00001              \\
CLEAN             &  &Y                     \\
CLEAN\_PARAM      &  &1                     \\
--------------------& Photometry &--------------------\\
PHOT\_APERTURES   &  &17.65,2.94,11.76,35        \\
PHOT\_AUTOPARAMS  &  &2.5, 3.5              \\
PHOT\_PETROPARAMS &  &2.0, 3.5              \\
PHOT\_AUTOAPERS   &  &0.0, 0.0              \\
PHOT\_FLUXFRAC    &  &0.2,0.5,0.8           \\
SATUR\_LEVEL      &  &50000                 \\
SATUR\_KEY        &  &SATURATE              \\
MAG\_ZEROPOINT    & SC NB816    & 31.07  \\
                  & SC i        & 31.17  \\
                  & HSC NB816   & 26.98  \\
                  & HSC i       & 27.02  \\
MAG\_GAMMA        &  &4                     \\
GAIN              &  &0                     \\
GAIN\_KEY         &  &GAIN                  \\
PIXEL\_SCALE      &  &0.17                  \\
SEEING\_FWHM      & SC NB816    & 0.95  \\
                  & SC i        & 0.95  \\
                  & HSC NB816   & 0.73  \\
                  & HSC i       & 0.66  \\
--------------------& Background &--------------------\\
BACK\_TYP         &  &AUTO                  \\
BACK\_VALUE       &  &0                     \\
BACK\_SIZE        &  &128                   \\
BACK\_FILTERSIZE  &  &3                     \\
BACKPHOTO\_TYPE   &  &LOCAL                 \\
BACKPHOTO\_THICK  &  &30                    \\

\hline 
\end{tabular}
}
\end{table}
\begin{table}[]
\centering
	\caption{Observation Dates of i'/i and NB816 bands with SC and HSC}
	\label{tab:observing_date}
	\setlength{\tabcolsep}{0.2cm}{
\begin{tabular}{c|cc} \hline \hline
filter & SC  & HSC \\ \hline
i'/i   & \begin{tabular}[c]{@{}c@{}} Feb. 2004, \\  Jan. 2004\end{tabular}  & \begin{tabular}[c]{@{}c@{}} Jan. 2020,\\  May. 2019,\\  Apr. 2017,\\  Mar. 2017,\\  Feb. 2017,\\  Jan. 2017,\\  Dec. 2016,\\  Nov. 2016,\\  Feb. 2016\end{tabular} \\ \hline
NB816  & \begin{tabular}[c]{@{}c@{}} Apr. 2005,\\  Mar. 2005,\\  Feb. 2005,\\  Apr. 2004\end{tabular} & \begin{tabular}[c]{@{}c@{}} Mar. 2019,\\  Feb. 2019,\\  Feb. 2016\end{tabular}    \\ \hline \hline                                                                                                            
\end{tabular}
}
\end{table}



\end{document}